\documentclass[
reprint,
superscriptaddress,
amsmath,amssymb,
aip,
jcp,
]{revtex4-2}

\usepackage{graphicx}
\usepackage{dcolumn}
\usepackage{bm}
\usepackage{extarrows}
\usepackage[colorlinks = true,
            linkcolor = blue,
            urlcolor  = blue,
            citecolor = blue,
            anchorcolor = blue]{hyperref}
\usepackage{makecell}

\begin{document}

\title[]{\large Electron transfer in confined electromagnetic fields: a unified Fermi's golden rule rate theory and extension to lossy cavities
\vspace{0.2cm}}

\author{Wenxiang Ying}
\email{wying3@sas.upenn.edu}
\affiliation{Department of Chemistry, University of Pennsylvania, Philadelphia, Pennsylvania 19104, USA}

\author{Abraham Nitzan}
\email{anitzan@sas.upenn.edu}
\affiliation{Department of Chemistry, University of Pennsylvania, Philadelphia, Pennsylvania 19104, USA}
\affiliation{School of Chemistry, Tel Aviv University, Tel Aviv 69978, Israel}

\begin{abstract}
With the rapid development of nanophotonics and cavity quantum electrodynamics, there has been growing interest in how confined electromagnetic fields modify fundamental molecular processes such as electron transfer. 
In this paper, we revisit the problem of nonadiabatic electron transfer (ET) in confined electromagnetic fields studied in [J. Chem. Phys. 150, 174122 (2019)] and present a unified rate theory based on Fermi’s golden rule (FGR). By employing a polaron-transformed Hamiltonian, we derive analytic expressions for the ET rate correlation functions that are valid across all temperature regimes and all cavity mode time scales. In the high-temperature limit, our formalism recovers the Marcus and Marcus–Jortner results, while in the low-temperature limit it reveals the emergence of the energy gap law. We further extend the theory to include cavity loss by using an effective Brownian oscillator spectral density, which enables closed-form expressions for the ET rate in lossy cavities. As applications, we demonstrate two key cavity-induced phenomena: (i) resonance effects, where the ET rate is strongly enhanced with certain cavity mode frequencies, and (ii) electron-transfer-induced photon emission, arising from the population of cavity photon Fock states during the ET process. These results establish a general framework for understanding how confined electromagnetic fields reshape charge transfer dynamics, and suggest novel opportunities for controlling and probing ET reactions in nanophotonic environments. 
\end{abstract}

\maketitle

\section{Introduction}
The possibility of harnessing quantum-electrodynamic (QED) effects to modify chemical reactions has recently drawn great attention. 
Recent experiments~\cite{Ebbesen_angew_2016, Ebbesen_science_2019, Simpkins2023, Ebbesen_nanophotonics_2020, Ebbesen_Angew_2021, Hirai_2020, verdelli_2024} have shown that vibrational strong coupling (VSC) can resonantly alter ground-state chemical reactivity, offering new strategies in synthetic chemistry. 
Meanwhile, studies in the electronic strong coupling (ESC) regime have demonstrated the potential to reshape nonadiabatic dynamics and photochemical reactions inside optical cavities~\cite{Zeng2023JACS, Lee2024JACS, Hutchinson2012ACIE, Munkhbat2018SA, Ng2015AOM, Mony2021AFM}. 
Among these processes, one of the most prominent examples is electron transfer (ET) in condensed phases~\cite{Marcus_1956, Hush_1958, Levich1966, Marcus_1985, Nitzan, Kuhn_2011}, which is ubiquitous in organic, inorganic, and biological systems alike. Cavity-modified ET dynamics thus provide unique opportunities to control charge transport and chemical reactivity in nanophotonic environments. 
An early theoretical study by Sch\"afer {\it et al.}~\cite{Schafer_PNAS2019} demonstrated that cavity QED effects can indeed alter the Dexter charge-transfer mechanism. 
Nevertheless, developing a general theoretical framework for hybrid matter–field systems remains a challenging task, as these processes involve a rich dynamical interplay among electronic, nuclear, and photonic degrees of freedom (DOF).

Since the theoretical work by Semenov and Nitzan~\cite{Nitzan_2019}, cavity-modified condensed phase ET processes have been extensively explored by theorists.
For example, Huo, {\it et al.}~\cite{Huo_2021} applied the ring polymer molecular dynamics (RPMD) approach, which uses ring polymer representation for the cavity photon mode to account for its quantum effects. 
Besides, they had also studied a senario of polariton-mediated electron transfer (PMET)~\cite{Arkajit_PMET_2020, Koessler2025CS}. 
Beratan, {\it et al.}~\cite{Beratan_JPCL2022} studied cavity modulated ET rates in donor–bridge–acceptor (DBA) systems. 
Meanwhile, Su, {\it et al.}~\cite{Su_2022} studied ET under a Floquet modulation in the DBA systems.
Wei and Hsu~\cite{Hsu_JPCL2022} developed a QED version of ET by incorporating a continuum of photon modes into the Marcus model Hamiltonian, laying an important foundation for the macroscopic QED descriptions~\cite{Hsu_JCP2019, Hsu_JCP2022-1, Hsu_JCP2022-2, Svendsen_2024, Hsu_JPCL2025}. 
Hayashi, {\it et al.}~\cite{Hayashi_JCP2024} studied the role of cavity strong coupling on ET rate at electrode–electrolyte interface.
The series of work done by Geva, {\it et al.}~\cite{Geva_JPCL2022, Geva_JPCC2023, Geva_JCP2023, Geva_JPCL2024, Geva_arXiv2025} had extensively explored cavity modified ET using Fermi's golden rule (FGR) and related linearized semicalssical approaches~\cite{Geva_JPCL2022}, addressing many-mode effects~\cite{Geva_JCP2023}, and even implementing simulations on noisy intermediate-scale quantum (NISQ) devices~\cite{Geva_JPCL2024}. 
In addition, a number of theoretical work have also examined the possibility of collective effects in cavity-modified ET~\cite{Herrera_Spano_PRL, Mauro_PRB2021, Wellnitz_JCP2021, Chen_JCP2024, Koessler2025CS}. 

Despite significant progress, existing studies still face important limitations. The widely used Marcus~\cite{Marcus_1956} and Marcus–Jortner~\cite{Jortner_1968} formulas~\footnote{We use ``Marcus-Jortner'' as a reference to the formulation of ET theory most relevant to our present extension. However, the seminal contributions of Levich and Hush to the general theory of electron transfer should be acknowledged. }, for instance, rely on the high-temperature approximation for nuclear vibrations; they generally fail in the low-temperature regime, where quantum effects become essential, and RPMD-based simulations might also become more challenging~\cite{Manolopoulos_2019}. The FGR calculations of Geva {\it et al.} treat both the vibronic environment and the quantized cavity mode on equal footing, and is applicable in all temperature regimes, but is generally restricted to the fast-cavity-mode limit (see Secion~\ref{sec:fast-cavity} for definition) and without including cavity loss~\cite{Geva_JPCL2022, Geva_JPCC2023, Geva_JCP2023, Geva_JPCL2024}. Thus, a general theoretical framework that is applicable to both fast and slow cavity modes, to high- and low-temperature regimes, and to both lossless and lossy cavities is warranted.

In this paper, we revisit the cavity-modified electron transfer problem for the single-molecule strong coupling case within the framework of FGR. In particular, an analytic correlation function has been derived, which generally works with fast or slow cavity modes, high or low temperature regimes, lossless or lossy cavities alike.
We show explicitly how FGR reduces to the Marcus and Marcus–Jortner expressions under appropriate limits, while also recovering the energy gap law~\cite{Jortner_MolPhys1970, Jortner_JLumi1970} at low temperatures. To account for realistic nanophotonic environments, we further incorporate a Brownian oscillator spectral density for the cavity modes, thereby generalizing the FGR rate to lossy cavities. As applications of this framework, we investigate (i) resonance effects, where the ET rate is selectively enhanced when the cavity frequency matches relevant energetic parameters, and (ii) ET-induced photon emission, arising from cavity population during the ET process. Together, these results provide a unified theory of cavity-modified ET and suggest new physical mechanisms for controlling and probing charge-transfer dynamics in confined electromagnetic fields.

This paper is organized as follows. In Section~\ref{sec:hams}, we introduce the model Hamiltonian. In Section~\ref{sec:FGR}, we discuss the FGR rate formulated in the time domain and its approximated cases. In Section~\ref{sec:loss}, we extend the FGR theory to lossy cavities. In Section~\ref{sec:numerical}, we present numerical results examining the implications of the theoretical expressions. Finally, we conclude in Section~\ref{sec:conclusion} and further discuss potential applications and limitations of our results.

\section{Model Hamiltonian} \label{sec:hams}
In this section, we express the model Hamiltonian in two different representations, {\it i.e.}, the Pauli-Fierz Hamiltonian under a linear vibronic coupling form, and the polaron transformed form, in preparation for the development of FGR rate theory. For simplicity, here we assume the long-wavelength approximation holds and consider only a single cavity mode with frequency $\omega$. 

\subsection{The linear vibronic coupling (LVC) form}
The Pauli-Fierz Hamiltonian can be derived via Power-Zienau-Woolley (PZW) gauge transformation~\cite{Power1959PTRSA, CohenTannoudji1997, Woolley1974JPBAMP} plus a constant phase shift~\cite{Arkajit_Chemrev_2023} based on the minimal coupling Hamiltonian under the Coulomb gauge. It is in a LVC form and is expressed as follows
\begin{widetext}
\begin{align} \label{eq:Hams-LVC}
    \hat{H} &= \Big(E_D + \frac{|g'_D|^2}{\hbar \omega} \Big) |\text{D}\rangle \langle \text{D}| + \Big(E_A + \frac{|g'_A|^2}{\hbar \omega} + \sum_j \frac{\lambda^2_j}{\hbar \nu_j} \Big) |\text{A}\rangle \langle \text{A}| + \left[H_{DA} + \frac{(g'_D + g'_A)t'_{DA}}{\hbar \omega} \right] |\text{D}\rangle \langle \text{A}| \notag\\
    &~~~ + \Big[H_{AD} + \frac{(g'_D + g'_A)t'_{AD}}{\hbar \omega} \Big] |\text{A}\rangle \langle \text{D}| + \sum_j \hbar \nu_j \hat{b}^\dagger_j \hat{b}_j + \hbar \omega \hat{a}^\dagger \hat{a} \\
    &~~~ + |\text{D}\rangle \langle \text{D}| \otimes g'_D (\hat{a} + \hat{a}^\dagger) + |\text{A}\rangle \langle \text{A}| \otimes [g'_A (\hat{a} + \hat{a}^\dagger) +  \sum_j \lambda_j (\hat{b}_j + \hat{b}^\dagger_j)] + |\text{D}\rangle \langle \text{A}| \otimes t'_{DA} (\hat{a} + \hat{a}^\dagger) + |\text{A}\rangle \langle \text{D}| \otimes t'_{AD} (\hat{a} + \hat{a}^\dagger). \notag
\end{align}
\end{widetext}
Details on the derivation for Eq.~\ref{eq:Hams-LVC} are provided in Appendix~\ref{apdx:Hams-original}.
In Eq.~\ref{eq:Hams-LVC}, $|\text{D}\rangle$ and $|\text{A}\rangle$ stand for states with the excess electron located on the donor and the acceptor sites, with corresponding electronic energy $E_D$ and $E_A$, respectively. $H_{DA}$ and $H_{AD}$ stand for the electron tunneling coupling between $|\text{D}\rangle$ and $|\text{A}\rangle$. $\hat{b}^\dagger_j$ ($\hat{b}_j$) denotes the creation (annihilation) operator of a nuclear vibrational (phonon) mode of frequency $\nu_j$ associated with the bath (inter/intramolecular vibration), with $\lambda_j$ the vibronic coupling parameter. Furthermore, $\hat{a}^\dagger$ ($\hat{a}$) denote the creation (annihilation) operator of the cavity mode, with $\omega$ the cavity mode frequency, $\{g'_D, g'_A\}$ and $\{t'_{DA}, t'_{AD}\}$ the diagonal and off-diagonal light-matter coupling strength, respectively (see their definitions in Appendix~\ref{apdx:Hams-original}). Note that they are different from the $\{g_D, g_A\}$ and $\{t_{DA}, t_{AD}\}$ parameters used in Ref.~\citenum{Nitzan_2019} (see also Eq.~\ref{eq:LMCS}). 

According to the Caldeira-Leggett model~\cite{Caldeira_Leggett_1983}, the phonon bath $\sum_j \hbar \nu_j \hat{b}^\dagger_j \hat{b}_j$ as well as its coupling with the reaction coordinate can be described by the spectral density as follows,
\begin{align}
    J_\text{vib}(\tilde{\omega}) = \frac{\pi}{\hbar} \sum_j \lambda^2_j \delta(\tilde{\omega} - \nu_j),
\end{align}
and the reorganization energy is
\begin{align}
    E_R = \frac{1}{\pi} \int_0^\infty d\tilde{\omega}~ \frac{J_\text{vib}(\tilde{\omega})}{\tilde{\omega}} = \sum_j \frac{\lambda^2_j}{\hbar \nu_j}. 
\end{align}

\subsection{The polaron transformed form}
The polaron transform (PT) for the LVC Hamiltonian in Eq.~\ref{eq:Hams-LVC} with respect to both the bath phonon and cavity photon DOF is defined as $\mathcal{\hat{H}} \equiv e^{\hat{S}} \hat{H} e^{- \hat{S}}$, with 
\begin{align} \label{eq:S-operator}
    \hat{S} &= |\text{D} \rangle \langle \text{D}| \otimes \frac{g'_D}{\hbar \omega} (\hat{a}^\dagger - \hat{a}) \notag\\
    &~~~ + |\text{A} \rangle \langle \text{A}| \otimes \Big[\frac{g'_A}{\hbar \omega} (\hat{a}^\dagger - \hat{a}) + \sum_j \frac{\lambda_j}{\hbar \nu_j} (\hat{b}^\dagger_j - \hat{b}_j) \Big].
\end{align}
Note that the projection operators $|\text{D} \rangle \langle \text{D}|$ and $|\text{A} \rangle \langle \text{A}|$ commute with each other, so one can linearly decompose $e^{\pm \hat{S}}$ arbitrarily without influencing the final result. As such, the first polaron transform (for the cavity mode, see also Eq.~18 in Ref.~\citenum{Nitzan_2019}) and the second polaron transform (for the bath phonon modes, see also Eq.~20 in Ref.~\citenum{Nitzan_2019}) commute with each other. The outcome does not depend on the order of acting the two polaron transform operators, thus does not face with the non-commuting operators problem as discussed in Ref.~\citenum{Segal_2023}. 
After PT, the Hamiltonian is expressed in a system-plus-bath form as
\begin{align} \label{eq:Hams-PT-3}
    \mathcal{\hat{H}} \equiv e^{\hat{S}} \hat{H} e^{- \hat{S}} = \mathcal{\hat{H}}_\text{S} + \hat{h}_\text{B} + \mathcal{\hat{H}}_\text{SB},
\end{align}
with purely diagonal system Hamiltonian $\mathcal{\hat{H}}_\text{S} = E_D |\text{D}\rangle \langle \text{D}| + E_A |\text{A}\rangle \langle \text{A}|$ and bath Hamiltonian $\hat{h}_\text{B} = \hbar \omega \hat{a}^\dagger \hat{a} + \sum_j \hbar \nu_j \hat{b}^\dagger_j \hat{b}_j$, and a purely off-diagonal system-bath coupling $\mathcal{\hat{H}}_\text{SB} = |\text{D}\rangle \langle \text{A}| \otimes \hat{F}_\text{DA} + |\text{A}\rangle \langle \text{D}| \otimes \hat{F}_\text{AD}$, where 
\begin{subequations}
\begin{align}
    \hat{F}_\text{DA} &= \left\{H_{DA} + t'_{DA} \Big[- g'_{DA} + (\hat{a} + \hat{a}^\dagger) \Big] \right\} \notag\\
    &~~~ \times e^{g'_{DA} (\hat{a}^\dagger - \hat{a}) - \sum_j \frac{\lambda_j}{\hbar \nu_j} (\hat{b}^\dagger_j - \hat{b}_j)}, \label{eq:FDA} \\
    \hat{F}_\text{AD} &= \left\{H_{AD} + t'_{AD} \Big[g'_{DA} + (\hat{a} + \hat{a}^\dagger) \Big] \right\} \notag\\
    &~~~ \times e^{- g'_{DA} (\hat{a}^\dagger - \hat{a}) + \sum_j \frac{\lambda_j}{\hbar \nu_j} (\hat{b}^\dagger_j - \hat{b}_j)}, \label{eq:FAD} 
\end{align}
\end{subequations}
and for simplicity we have defined
\begin{equation} \label{eq:gDA}
    g'_{DA} \equiv \frac{g'_D - g'_A}{\hbar \omega}.
\end{equation}
The derivation for the polaron transformed Hamiltonian is presented in Supplementary Material, Section I.

\section{Fermi's Golden Rule (FGR) Rate Theory} \label{sec:FGR}
The FGR rate theory describes electron transfer on harmonic potential energy surfaces, has long served as a standard framework for nonadiabatic charge-transfer dynamics.
In this section, we briefly introduce the FGR rate theory formulated in the time-domain, and recover the Marcus and Marcus-Jortner theories under the high-temperature limit for the nuclei. We also discuss the emergence of the energy gap law~\cite{Jortner_MolPhys1970, Jortner_JLumi1970} under the low-temperature limit. 

Based on the PT Hamiltonian in Eq.~\ref{eq:Hams-PT-3}, the transition rate between the donor and the acceptor states can be expressed as~\cite{Nitzan}
\begin{align} \label{eq:FGR-LRT}
    k_{\text{D}\to \text{A}} = \frac{1}{\hbar^2} \int_{-\infty}^{\infty} dt~ e^{- i \Delta G_0 t/\hbar} C_{ff}(t),
\end{align}
where $- \Delta G_0 = E_D - E_A$ is the donor-acceptor energy gap, and the force-force correlation function is defined as
\begin{align} \label{eq:FFCF}
    C_{ff}(t) \equiv \text{Tr} [e^{i\hat{h}_\text{B}t / \hbar} \hat{F}_\text{DA} e^{- i\hat{h}_\text{B}t / \hbar} \hat{F}_\text{AD} \hat{\rho}^\text{eq}_\text{B}],
\end{align}
where $\hat{F}_\text{DA}$ and $\hat{F}_\text{AD}$ are the bath coupling terms in Eq.~\ref{eq:FDA} and \ref{eq:FAD}, respectively, and $\hat{\rho}^\text{eq}_\text{B} \equiv e^{-\beta\hat{h}_\text{B}} / \text{Tr}[e^{-\beta\hat{h}_\text{B}}]$ is the bath thermal density matrix, with the reciprocal temperature $\beta \equiv 1 / (k_\text{B} T)$, $k_\text{B}$ is the Boltzmann constant, $T$ is the temperature. 

Note that there are alternative forms of the FGR rate expression apart from Eq.~\ref{eq:FGR-LRT}, for example, the one discussed by Geva, {\it et al.}~\cite{Sun_Geva_2016, Sun_Geva_2016_2, Geva_JCP2023, Geva_JPCL2022, Geva_JPCC2023}. These forms are all equivalent as is shown in Appendix~\ref{apdx:FGR-2}. 

In particular, the correlation function defined in Eq.~\ref{eq:FFCF} can be evaluated analytically, 
\begin{align} \label{eq:FFCF-2}
    C_{ff}(t) &= [h(t) + g(t)] \cdot e^{f(t)},
\end{align}
where
\begin{widetext}
\begin{subequations} \label{eq:hgf_t}
\begin{align}
    h(t) &= \Big\{ H_{DA} +  t'_{DA} g'_{DA} \Big[- \cos(\omega t) + i \sin(\omega t) \coth(\frac{\beta \hbar \omega}{2}) \Big] \Big\} \times \Big\{ H_{AD} + t'_{AD} g'_{DA} \Big[- \cos(\omega t) + i \sin(\omega t) \coth(\frac{\beta \hbar \omega}{2}) \Big] \Big\}, \\
    g(t) &= t'_{DA} t'_{AD} \Big\{ \cos(\omega t) \coth(\frac{\beta \hbar \omega}{2}) - i \sin(\omega t)  \Big\}, \\
    f(t) &= - |g'_{DA}|^2 \Big\{ [1 - \cos(\omega t)] \coth(\frac{\beta \hbar \omega}{2}) + i \sin(\omega t) \Big\} - \sum_j \frac{\lambda^2_j}{(\hbar \nu_j)^2} \Big\{ [1 - \cos(\nu_j t)] \coth(\frac{\beta \hbar \nu_j}{2}) + i \sin(\nu_j t) \Big\}. 
\end{align}
\end{subequations}
\end{widetext}
Derivation for Eqs.~\ref{eq:FFCF-2}-\ref{eq:hgf_t} are presented in Supplementary Material, Section II, being a general expression that cover all temperature regimes and cavity time scales, which also serves as the basis to further apply various approximations. 

It is well-known that for conventional ET problems (without cavity coupling), under the high-temperature limit for bath phonon modes, such that $\sum_j \frac{\lambda_j}{\hbar \nu_j} \overline{n}_j \gg 1$ with $\overline{n}_j = 1 / (e^{\beta \hbar \nu_j} - 1)$, the FGR rate expression reduces to the Marcus theory~\cite{Marcus_1956, Nitzan}, 
\begin{align} \label{eq:Marcus}
    k_{\text{D}\to \text{A}} &= H_{DA} H_{AD} \cdot \sqrt{\frac{\pi \beta}{\hbar^2 E_R}} \cdot e^{- \beta \frac{[- \Delta G_0 - E_R]^2}{4 E_R}}.
\end{align}
In the following, we discuss the way coupling to the cavity mode influences this rate in different regimes of temperatures and time scales. 

\subsection{fast cavity mode, slow electron tunneling} \label{sec:fast-cavity}
Consider first the case where the cavity mode is fast relative to the electron tunneling time scale~\footnote{This is a mathematically valid but physically an unlikely limit, given the electron tunneling timescale, essentially the time the electron spends under the barrier, is of the same order of molecular electronic transitions ($\sim$1 eV or above, with sub-fs time scale).}. In this case, we can use Born-Oppenheimer approximation with the molecular electronic and nuclear DOF assumed slow compared to the cavity dynamics. 
In other words, the reorganization of the cavity mode DOF during electron tunneling process is almost instantaneous, meaning that one can take $g'_{DA} \approx$ 0 (Eq.~\ref{eq:gDA}). 
As a result, the force-force correlation function in Eq.~\ref{eq:FFCF-2} can be simplified as
\begin{align} \label{eq:FFCF-2A}
    &C_{ff}(t) \notag\\
    &= \Big\{ H_{DA} H_{AD} + t'_{DA} t'_{AD} \big[ \cos(\omega t) \coth(\frac{\beta \hbar \omega}{2}) - i \sin(\omega t) \big] \Big\} \notag\\
    &~~~\times e^{- \sum_j \frac{\lambda^2_j}{(\hbar \nu_j)^2} \Big\{ [1 - \cos(\nu_j t)] \coth(\frac{\beta \hbar \nu_j}{2}) + i \sin(\nu_j t) \Big\}}, 
\end{align}
which is exactly the FGR correlation function derived by Geva, {\it et al}. in Ref.~\citenum{Geva_JPCC2023}. 
Below, we further consider several special limits of the temperature. 

Consider first the (usually unphysical for the fast cavity case) high-temperature limit when $k_\text{B} T \gg \hbar \omega, \hbar \nu_j$. 
In this case, the cavity mode $\delta \hat{q} = \hat{a} + \hat{a}^\dagger$ can be regarded as a (classical) fluctuating bridge, since $\langle \delta \hat{q} \rangle = 0$, while 
\begin{equation}
    \langle (\delta \hat{q})^2 \rangle = \frac{1 + e^{-\beta \hbar \omega}}{1 - e^{-\beta \hbar \omega}} ~~~\xrightarrow[]{\beta \to 0}~~~=\frac{2}{\beta \hbar \omega} > 0. 
\end{equation}
Here, the thermal average with respect to the cavity mode DOF is $\langle \cdot \rangle = \sum_{n=0}^\infty \langle n| \cdot |n\rangle e^{-(n+1/2) \beta \hbar \omega} / \mathcal{Z}$, with the partition function $\mathcal{Z} = e^{-\beta \hbar \omega / 2} / (1 - e^{-\beta \hbar \omega})$.
The cavity mode coupling leads to a rate expression that contains a series of additional terms~\cite{Troisi_2003} associated to the original Marcus rate in Eq.~\ref{eq:Marcus}. 
As $\beta \to 0$, Eq.~\ref{eq:FFCF-2A} leads to a Marcus-type ET rate expressed as following~\cite{Goychuk_1995, Medvedev_1997, Troisi_2003},
\begin{align} \label{eq:Marcus-A}
    k_{\text{D}\to \text{A}} &= (k^\text{(0)} + k^\text{(1)} + k^\text{(2)} + \mathcal{O}(\beta^2) ) \notag\\
    &~~~ \times \sqrt{\frac{\pi \beta}{\hbar^2 E_R}} \cdot e^{- \beta \frac{[- \Delta G_0 - E_R]^2}{4 E_R}}.
\end{align}
where the first few lowest order contributions from the fluctuating bridge explicitly read as
\begin{subequations}
\begin{align}
    k^\text{(0)} &= \frac{2 t'_{DA} t'_{AD}}{\beta \hbar \omega}, \label{eq:k_0_A} \\
    k^\text{(1)} &= H_{DA} H_{AD} - t'_{DA} t'_{AD} \frac{\hbar \omega}{2E_R}, \\
    k^\text{(2)} &= - t'_{DA} t'_{AD} \frac{\beta \hbar \omega}{4E_R} \Big[2 (\Delta G_0 + E_R) + \frac{(\Delta G_0 + E_R)^2}{E_R} \Big].
\end{align}
\end{subequations}
One sees from Eq.~\ref{eq:Marcus-A} that the cavity mode provides additional tunneling channel due to thermal fluctuations, with the leading order $k^\text{(0)}$ (Eq.~\ref{eq:k_0_A}) magnitude proportional to $\beta^{-1}$, the first-order correction $k^\text{(1)} \propto \beta^0$, the second-order correction $k^\text{(2)} \propto \beta$, and $\mathcal{O}(\beta^2)$ denotes small residual terms with order of $\beta^2$ or higher. 

Next, we consider a moderate temperature regime that satisfies the high-temperature limit for bath phonon modes while low-temperature limit for the cavity mode, {\it i.e.}, $\hbar \omega \gg k_\text{B} T \gg \hbar \nu_j$. 
Under this circumstance, the high frequency cavity mode further adds quantum mechanical corrections to the classical nuclei in Marcus theory, such that the Marcus theory is generalized to the Marcus-Jortner (MJ) theory~\cite{Jortner_1968, Nitzan}. 
Specifically, Eq.~\ref{eq:FFCF-2A} leads to the following MJ rate expression,
\begin{align} \label{eq:MJ_rate_A}
    k_{\text{D}\to \text{A}} &= \sqrt{\frac{\pi \beta}{\hbar^2 E_R}} \Big\{ H_{DA} H_{AD} \cdot e^{- \beta \frac{[- \Delta G_0 - E_R]^2}{4 E_R}} \notag\\
    &~~~ + t'_{DA} t'_{AD} \cdot e^{- \beta \frac{[- \Delta G_0 - E_R - \hbar \omega]^2}{4 E_R}} \Big\}.
\end{align}
Eq.~\ref{eq:MJ_rate_A} is just Eq.~38 of Ref.~\citenum{Nitzan_2019}. 
Note that Eq.~\ref{eq:MJ_rate_A} has also been derived by Geva, {\it et al}.~\cite{Geva_JPCC2023} via reduction from FGR. 

Finally, consider the low-temperature limit when $\hbar \omega, \hbar \nu_j \gg k_\text{B} T$, so that the Marcus / MJ theory breaks down. 
Eq.~\ref{eq:FFCF-2A} leads to the following rate expression,
\begin{align} \label{eq:low-T-A}
    &k_{\text{D}\to \text{A}} = \frac{2\pi}{\hbar^2} \cdot e^{- \sum_j \overline{\lambda}^2_j} \cdot \sum_{\{m_j\}} \Big\{ H_{DA} H_{AD} \cdot \delta(\omega_{21} - \sum_j m_j \nu_j) \notag\\
    & + t'_{DA} t'_{AD} \cdot \delta(\omega_{21} - \omega - \sum_j m_j \nu_j) \Big\} \cdot \prod_j \frac{\overline{\lambda}_j^{2m_j}}{m_j!}, 
\end{align}
where $\hbar \omega_{21} = E_D - E_A$ is the energy gap, and for simplicity we have denoted $\overline{\lambda}_j = \lambda_j / (\hbar \nu_j)$. 
We use Eq.~\ref{eq:low-T-A} to analyze the scaling relation between the ET rate and the donor-acceptor energy gap, which manifests as the energy gap law (EGL)~\cite{Jortner_MolPhys1970, Jortner_JLumi1970}. 
Here, we consider the parameter regime with a very large energy gap ($\omega_{21} \gg \nu_j$) and under the weak coupling limit ($\overline{\lambda}_j \ll 1$), so that the summation with respect to $m_j$ is dominated by the terms with the smallest excitation number $m_j$, which in turn corresponds to the highest mode frequency~\cite{Nitzan}. 
Eq.~\ref{eq:low-T-A} reduces to outside the cavity case if further taking $t'_{DA} = 0$, where the energy matching condition in the first term $\delta(\omega_{21} - \sum_j m_j \nu_j)$ dominates the EGL~\cite{Nitzan}. To be specific,
\begin{align} \label{eq:EGL_A_1}
    k_{\text{D}\to \text{A}} (\omega_{21}) \sim \exp\Big(\frac{\omega_{21}}{\omega_\text{c}}\ln \overline{\lambda}_\text{c}^2 - \frac{\omega_{21}}{\omega_\text{c}} \ln \frac{\omega_{21}}{\omega_\text{c}} \Big), 
\end{align}
where $\omega_\text{c}$ is the bath characteristic phonon frequency, and $\overline{\lambda}_\text{c}$ is its rescaled coupling strength to the electronic state. One sees that as $\overline{\lambda}_\text{c} < 1$, $k_{\text{D}\to \text{A}} (\omega_{21})$ decays exponentially or faster with respect to $\omega_{21}$. 

With the presence of cavity mode coupling, the second term $t'_{DA} t'_{AD} \cdot \delta(\omega_{21} - \omega - \sum_j m_j \nu_j)$ also contributes to the ET rate, which gives rise to the following EGL,
\begin{align} \label{eq:EGL_A_2}
    k_{\text{D}\to \text{A}} (\omega_{21}) \sim \exp\Big(\frac{\omega_{21} - \omega}{\omega_\text{c}}\ln \overline{\lambda}_\text{c}^2 - \frac{\omega_{21} - \omega}{\omega_\text{c}} \ln \frac{\omega_{21} - \omega}{\omega_\text{c}} \Big). 
\end{align}
One sees that Eq.~\ref{eq:EGL_A_2} only shifts Eq.~\ref{eq:EGL_A_1} by an amount of $\omega$, but keeps the same scaling relation (given the same $\omega_\text{c}$ and $\overline{\lambda}_\text{c}$). This will lead to a same slope in the $k_{\text{D}\to \text{A}}$ v.s. $-\Delta G_0$ diagram for outside / inside cavity cases. 

\subsection{slow cavity mode, fast electron tunneling} \label{sec:slow-cavity}
We next examine the situation in which electron tunneling occurs on a timescale much shorter than that of the cavity mode.
In this regime, the tunneling rate is determined solely by the initial configuration of the cavity field, which, similar to nuclear coordinates, can be regarded as frozen during the tunneling event.
Accordingly, the influence of the cavity mode on electron transfer resembles that of other slow nuclear or environmental modes, which contributes negligibly to electron tunneling ({\it i.e.}, $t'_{DA}$, $t'_{AD} \ll H_{DA}$, $H_{AD}$) but mainly to bath modes reorganization, thus one can approximately take $t'_{DA} = t'_{AD} \approx 0$. 
As a result, the force-force correlation function in Eq.~\ref{eq:FFCF-2} can be simplified as
\begin{align} \label{eq:FFCF-2B}
    &C_{ff}(t) = H_{DA} H_{AD} \cdot e^{- |g'_{DA}|^2 \Big\{ [1 - \cos(\omega t)] \coth(\frac{\beta \hbar \omega}{2}) + i \sin(\omega t) \Big\}} \notag\\
    &~~~\times e^{- \sum_j \frac{\lambda^2_j}{(\hbar \nu_j)^2} \Big\{ [1 - \cos(\nu_j t)] \coth(\frac{\beta \hbar \nu_j}{2}) + i \sin(\nu_j t) \Big\}}. 
\end{align}
Similar as Sec.~\ref{sec:fast-cavity}, we discuss several special limits of the temperature. 

In the high-temperature limit when $k_\text{B} T \gg \hbar \omega, \hbar \nu_j$. Eq.~\ref{eq:FFCF-2B} leads to the following Marcus-type rate expression,
\begin{align} \label{eq:Marcus-B}
    k_{\text{D}\to \text{A}} &= H_{DA} H_{AD} \cdot \sqrt{\frac{\pi \beta}{\hbar^2 \tilde{E}_R}} \cdot \exp \Big(- \beta \frac{[- \Delta G_0 - \tilde{E}_R]^2}{4 \tilde{E}_R} \Big),
\end{align}
where the modified reorganization energy $\tilde{E}_R = E_R + |g'_{DA}|^{2} \hbar \omega$. One sees that the cavity mode acts as classical nuclei and only contributes to the reorganization energy, such that $E_R \to \tilde{E}_R = E_R + |g'_{DA}|^{2} \hbar \omega$. 

In the moderate temperature regime that satisfies the high-temperature limit for bath phonon modes while low-temperature limit for the cavity mode, {\it i.e.}, $\hbar \omega \gg k_\text{B} T \gg \hbar \nu_j$. 
Eq.~\ref{eq:FFCF-2B} leads to the following MJ rate expression,
\begin{align} \label{eq:MJ_rate_B}
    k_{\text{D}\to \text{A}} &= \sqrt{\frac{\pi \beta}{\hbar^2 E_R}} \cdot H_{DA} H_{AD} \cdot e^{- |g'_{DA}|^2} \notag\\
    &~~~ \times \sum_{m=0}^\infty \frac{|g'_{DA}|^{2m}}{m!} \cdot e^{- \beta \frac{[- \Delta G_0 - E_R - m \hbar \omega]^2}{4 E_R}},
\end{align}
which reproduces Eq.~39 of Ref.~\citenum{Nitzan_2019} if one keeps the $t'_{DA} t'_{AD}$ term (contribution to electron tunneling) -- which is negligible under the condition of $\hbar \omega \gg k_\text{B} T$. See details in Supplementary Material, Section V. 

Eq.~\ref{eq:MJ_rate_B} contains an infinite sum of terms. One can alternatively express Eq.~\ref{eq:MJ_rate_B} in terms of convolution between a complex function and a Gaussian as follows~\cite{Jortner_1974, Jortner_1975, Jortner_1976, Cui_2021}, 
\begin{align} \label{eq:MJ_rate_conv-B}
    &k_{\text{D}\to \text{A}} = \int_{-\infty}^{\infty} d\tilde{\omega}~ G_\text{cav} (\tilde{\omega}) \cdot G_\text{ph}\Big(\frac{\Delta G_0}{\hbar} - \tilde{\omega} \Big), 
\end{align}
where
\begin{subequations}
\begin{align}
    &G_\text{cav} (\tilde{\omega}) = H_{DA} H_{AD} \cdot \int_{-\infty}^{\infty} dt~ e^{- i \tilde{\omega} t} \cdot e^{|g'_{DA}|^2 (e^{-i\omega t} - 1)}, \\
    &G_\text{ph}\Big(\frac{\Delta G_0}{\hbar} - \tilde{\omega} \Big) = \sqrt{\frac{\pi \beta}{\hbar^2 E_R}} e^{- \beta \frac{(- \Delta G_0 - E_R + \hbar \tilde{\omega})^2}{4E_R}}. \label{eq:Gph_w}
\end{align}
\end{subequations}
Eq.~\ref{eq:Gph_w} is a Gaussian. Eq.~\ref{eq:MJ_rate_conv-B} provides an alternative approaches to numerically evaluate the MJ rate (that avoids infinite sum). 

In the low-temperature limit when $\hbar \omega, \hbar \nu_j \gg k_\text{B} T$. Eq.~\ref{eq:FFCF-2B} leads to the following rate expression,
\begin{align} \label{eq:low-T-B}
    &k_{\text{D}\to \text{A}} = \frac{2\pi}{\hbar^2} \cdot e^{- |g'_{DA}|^2 - \sum_j \overline{\lambda}^2_j} \sum_{n} \sum_{\{m_j\}} H_{DA} H_{AD} \notag\\
    &\times \delta(\omega_{21} - n\omega - \sum_j m_j \nu_j) \cdot \frac{|g'_{DA}|^{2n}}{n!} \cdot \prod_j \frac{\overline{\lambda}_j^{2m_j}}{m_j!}, 
\end{align}
with $\overline{\lambda}_j = \lambda_j / (\hbar \nu_j)$. We consider the parameter regime with a very large energy gap ($\omega_{21} \gg \omega, \nu_j$) and under the weak coupling limit ($g'_{DA}, \overline{\lambda}_j \ll 1$), so that the summation with respect to $n$ and $m_j$ is dominated by the terms with the smallest excitation number $n$ or $m_j$, which in turn corresponds to the highest mode frequency~\cite{Nitzan}. 
Here, if we focus on a regime that the cavity mode frequency much larger than the bath phonon characteristic frequency ($\omega \gg \omega_\text{c}$), then the summation should be dominated by $n$. For an extreme case, we take $\overline{n} = \omega_{21} / \omega$ and $m_j = 0$ for all $j$. As such, the EGL reads as
\begin{align} \label{eq:EGL_B_1}
    k_{\text{D}\to \text{A}} (\omega_{21}) \sim \exp\Big(\frac{\omega_{21}}{\omega}\ln |g'_{DA}|^2 - \frac{\omega_{21}}{\omega} \ln \frac{\omega_{21}}{\omega} \Big), 
\end{align}
being different from the EGL scaling relation outside the cavity (Eq.~\ref{eq:EGL_A_1}), provided that $\{\omega, g'_{DA}\}$ differs from $\{\omega_\text{c}, \overline{\lambda}_\text{c}\}$. This will lead to different slopes in the $k_{\text{D}\to \text{A}}$ v.s. $-\Delta G_0$ diagram for outside / inside cavity cases. 


\subsection{Intermediate regime}
Finally, we consider the case when the electron tunneling has comparable time scale as the cavity mode. In this regime, there will be a dynamical interplay between the cavity mode and electron tunneling, thus one can neglect neither $\{t'_{DA}, t'_{AD}\}$ nor $g'_{DA}$, and cross terms between them will show up. 
Similar as Sec.~\ref{sec:fast-cavity}, we discuss several special limits of the temperature. 

In the high-temperature limit when $k_\text{B} T \gg \hbar \omega, \hbar \nu_j$. Eq.~\ref{eq:FGR-LRT} leads to the following Marcus-type rate expression, 
\begin{align} \label{eq:Marcus-C}
    k_{\text{D}\to \text{A}} &= (k^\text{(0)} + k^\text{(1)} + k^\text{(2)} + \cdots) \times \sqrt{\frac{\pi \beta}{\hbar^2 \tilde{E}_R}} e^{- \beta \frac{(\Delta G_0 + \tilde{E}_R)^2}{4 \tilde{E}_R}},
\end{align}
where $\tilde{E}_R = E_R + |g'_{DA}|^{2} \hbar \omega$. Here, the first few lowest order (with respect to $\beta$) contribution from the fluctuating bridge (cavity mode) explicitly reads as
\begin{widetext}
\begin{subequations}
\begin{align}
    k^\text{(0)} &= \frac{2 t'_{DA} t'_{AD}}{\beta \hbar \omega} - \frac{2 t'_{DA} t'_{AD} |g'_{DA}|^2}{\beta \tilde{E}_R}, \label{eq:k_0_C} \\
    k^\text{(1)} &= H_{DA} H_{AD} + \frac{\Delta G_0}{\tilde{E}_R} \Big( H_{DA} t'_{AD} + H_{AD} t'_{DA} \Big) g'_{DA} - t'_{DA} t'_{AD} \Big[ - |g'_{DA}|^2 \Big(\frac{\Delta G_0}{\tilde{E}_R} \Big)^2 + \frac{\hbar \omega}{2 \tilde{E}_R} \Big], \\
    k^\text{(2)} &= \frac{\beta \hbar \omega}{4 \tilde{E}_R} \Big[ \Big( H_{DA} t'_{AD} + H_{AD} t'_{DA} \Big) g'_{DA} \hbar \omega - t'_{DA} t'_{AD} \Big( 2 (\Delta G_0 + \tilde{E}_R - |g'_{DA}|^2 \hbar \omega) + \frac{(\Delta G_0 + \tilde{E}_R)^2}{\tilde{E}_R} \Big) \Big].
\end{align}
\end{subequations}
\end{widetext}
The zeroth order term $k^\text{(0)} \propto \beta^{-1}$, first order term $k^\text{(1)} \propto \beta^0$, second order term $k^\text{(2)} \propto \beta$, and so on. It is straightforward to see that Eq.~\ref{eq:Marcus-C} reduces to Eq.~\ref{eq:Marcus-A} by taking $g'_{DA} = 0$, and $\tilde{E}_R \to E_R$. On the other hand, Eq.~\ref{eq:Marcus-C} reduces to Eq.~\ref{eq:Marcus-B} by taking $t'_{DA} = t'_{AD} = 0$.

In the moderate temperature regime that satisfies the high-temperature limit for bath phonon modes while low-temperature limit for the cavity mode, {\it i.e.}, $\hbar \omega \gg k_\text{B} T \gg \hbar \nu_j$. Eq.~\ref{eq:FGR-LRT} leads to the following MJ rate expression, 
\begin{widetext}
\begin{align} \label{eq:MJ_rate-C}
    k_{\text{D}\to \text{A}} &= \sqrt{\frac{\pi \beta}{\hbar^2 E_R}} \cdot e^{- |g'_{DA}|^2} \sum_{m=0}^\infty \frac{|g'_{DA}|^{2m}}{m!} \Big\{ H_{DA} H_{AD} \cdot e^{- \beta \frac{[- \Delta G_0 - E_R - m \hbar \omega]^2}{4 E_R}} \\
    & + (t'_{DA} t'_{AD} - H_{DA} t'_{AD} g'_{DA} - H_{AD} t'_{DA} g'_{DA}) \cdot e^{- \beta \frac{[- \Delta G_0 - E_R - (m + 1) \hbar \omega]^2}{4 E_R}} + t'_{DA} t'_{AD} |g'_{DA}|^2 \cdot e^{- \beta \frac{[- \Delta G_0 - E_R - (m + 2) \hbar \omega]^2}{4 E_R}} \Big\}. \notag
\end{align}
\end{widetext}
It is straightforward to see that Eq.~\ref{eq:MJ_rate-C} reduces to Eq.~\ref{eq:MJ_rate_A} by taking $g'_{DA} = 0$, so that only the $m = 0$ term survives. On the other hand, Eq.~\ref{eq:MJ_rate-C} reduces to Eq.~\ref{eq:MJ_rate_B} by taking $t'_{DA} = t'_{AD} = 0$. 

Eq.~\ref{eq:MJ_rate-C} contains an infinite sum of terms. One can alternatively express it in terms of convolution between a complex function and a Gaussian as follows~\cite{Jortner_1974, Jortner_1975, Jortner_1976, Cui_2021}, 
\begin{align} \label{eq:MJ_rate_conv-C}
    &k_{\text{D}\to \text{A}} = \int_{-\infty}^{\infty} d\tilde{\omega}~ G_\text{cav} (\tilde{\omega}) \cdot G_\text{ph}\Big(\frac{\Delta G_0}{\hbar} - \tilde{\omega} \Big), 
\end{align}
where
\begin{subequations}
\begin{align}
    &G_\text{cav} (\tilde{\omega}) = \int_{-\infty}^{\infty} dt~ e^{- i \tilde{\omega} t} \Big\{ \Big[ H_{DA} - t'_{DA} g'_{DA} e^{-i\omega t} \Big] \notag\\
    &~~~ \times \Big[ H_{AD} - t'_{AD} g'_{DA} e^{-i\omega t} \Big] + t'_{DA} t'_{AD} e^{-i\omega t} \Big\} \notag\\
    &~~~ \times e^{|g'_{DA}|^2 (e^{-i\omega t} - 1)}, \\
    &G_\text{ph}\Big(\frac{\Delta G_0}{\hbar} - \tilde{\omega} \Big) = \sqrt{\frac{\pi \beta}{\hbar^2 E_R}} e^{- \beta \frac{(- \Delta G_0 - E_R + \hbar \tilde{\omega})^2}{4E_R}}, 
\end{align}
\end{subequations}
being an alternative approach for numerical evaluation of the MJ rate (that avoids infinite sum). It is straightforward to see that Eq.~\ref{eq:MJ_rate_conv-C} reduces to Eq.~\ref{eq:MJ_rate_conv-B} by taking $t'_{DA} = t'_{AD} = 0$. 

\begin{table*}[htbp]
\renewcommand{\arraystretch}{3.5}
\caption{An overview of approximations based on the FGR rate expression, and the corresponding physical picture of the cavity mode.} \vspace{0.1 cm}
\begin{tabular*}{\textwidth}{@{\extracolsep{\fill}} c | c | c | c }
\hline\hline
\makecell[c]{Approximations \vspace{0.4em}} & \makecell[c]{The thermal activation \\
/potential surface-crossing limit~~ \\
$k_\text{B}T \gg \hbar\omega, \hbar\nu_j$ \vspace{0.4em} } & \makecell[c]{Quantum cavity mode correction~~ \\
to classical nuclei \\
$\hbar\omega \gg k_\text{B}T \gg \hbar\nu_j$ \vspace{0.4em} } & \makecell[c]{The low temperature \\
and weak coupling limit \\
$\hbar\omega, \hbar\nu_j \gg k_\text{B}T$ \vspace{0.4em} } \\ [0.5ex]
\hline 
\makecell[c]{fast cavity mode \\
\& slow electron tunneling~ \\ 
$\{t'_{DA}, t'_{AD}\}$; $g'_{DA} = 0$ \vspace{0.5em} } & \makecell[c]{Cavity mode as a fluctuating bridge; \\ Eq.~\ref{eq:Marcus-A} \vspace{0.2em}} & \makecell[c]{Eq.~\ref{eq:MJ_rate_A} \vspace{0.2em}} & \makecell[c]{Eq.~\ref{eq:low-T-A} \vspace{0.2em}} \\ [0.5ex] \hline 
\makecell[c]{slow cavity mode \\
\& fast electron tunneling~ \\ 
$\{t'_{DA}, t'_{AD}\} = 0$; $g'_{DA}$ \vspace{0.5em} } & \makecell[c]{Cavity mode as classical nuclei and \\
contributes to reorganization energy; \\ Eq.~\ref{eq:Marcus-B} \vspace{0.2em}} & \makecell[c]{Eq.~\ref{eq:MJ_rate_B} or \ref{eq:MJ_rate_conv-B} \vspace{0.2em}} & \makecell[c]{Eq.~\ref{eq:low-T-B} \vspace{0.2em}} \\ [0.5ex] \hline 
\makecell[c]{intermediate regime \\ 
$\{t'_{DA}, t'_{AD}\}$; $g'_{DA}$ \vspace{0.6em} } & \makecell[c]{Eq.~\ref{eq:Marcus-C} \vspace{0.3em}} & \makecell[c]{Eq.~\ref{eq:MJ_rate-C} or \ref{eq:MJ_rate_conv-C} \vspace{0.3em}} & \makecell[c]{Eq.~\ref{eq:low-T-C} \vspace{0.3em}} \\ [0.5ex]
\hline\hline
\end{tabular*}
\label{tab:overview}
\end{table*}

In the low-temperature limit when $\hbar \omega, \hbar \nu_j \gg k_\text{B} T$. Eq.~\ref{eq:FGR-LRT} leads to the following rate expression, 
\begin{widetext}
\begin{align} \label{eq:low-T-C}
    k_{\text{D}\to \text{A}} &= \frac{2\pi}{\hbar^2} \cdot e^{- |g'_{DA}|^2 - \sum_j \overline{\lambda}^2_j} \cdot \sum_{n} \sum_{\{m_j\}} \Big\{ H_{DA} H_{AD} \cdot \delta(\omega_{21} - n\omega - \sum_j m_j \nu_j) \notag\\
    &~~~ + (t'_{DA} t'_{AD} - H_{DA} t'_{AD} g'_{DA} - H_{AD} t'_{DA} g'_{DA}) \cdot \delta(\omega_{21} - (n+1)\omega - \sum_j m_j \nu_j) \notag\\
    &~~~ + t'_{DA} t'_{AD} |g'_{DA}|^2 \cdot \delta(\omega_{21} - (n+2)\omega - \sum_j m_j \nu_j) \Big\} \cdot \frac{|g'_{DA}|^{2n}}{n!} \cdot \prod_j \frac{\overline{\lambda}_j^{2m_j}}{m_j!}, 
\end{align}
\end{widetext}
with $\overline{\lambda}_j = \lambda_j / (\hbar \nu_j)$. It is straightforward to see that Eq.~\ref{eq:low-T-C} reduces to Eq.~\ref{eq:low-T-A} by taking $g'_{DA} = 0$, so that only the $n = 0$ term survives. On the other hand, Eq.~\ref{eq:low-T-C} reduces to Eq.~\ref{eq:low-T-B} by taking $t'_{DA} = t'_{AD} = 0$. 
One can accordingly analyze the EGL scaling relations using Eq.~\ref{eq:low-T-C}. 

The derivation for Eqs.~\ref{eq:Marcus-C}, \ref{eq:MJ_rate-C}, and \ref{eq:low-T-C} are provided in Supplementary Material, Section III.  
For the sake of clarity, we also present in Table~\ref{tab:overview} an overview of the approximations discussed in this section, as well as the resulting rate expressions and physical picture of the cavity mode. 
We also note that there are still a variety of parameter regimes that are not covered by the limits discussed above, where we have implicitly assumed the cavity mode frequency is always much higher than the molecular vibration frequency. But this assumption can be violated since the highest molecular vibrations are about 0.4 eV. 

To conclude this section we note that our discussions above assumes that the FGR description holds. The assumption will break down under the electronic strong coupling regime where the adiabaticity increases and polaritons become the true physical states, leading to an interesting strong coupling scenario that are beyond FGR description. 

\section{Incorporation of Cavity Loss} \label{sec:loss}
In real experiments, the lifetime of the cavity mode $\tau_\mathrm{c}$ is finite due to its coupling with the far-field photon modes outside the cavity, which causes broadenings to the cavity mode spectrum. In this section, we discuss ET dynamics within lossy cavities and its associated rate theories. 

The interactions between the cavity mode and the far-field modes can also be described using a system-bath model, which is also known as the Gardiner-Collett Hamiltonian~\cite{Dutra2000PRA, Dutra2000JOB, Dutra2005JWS}. The total Hamiltonian is then expressed as
\begin{align} \label{eq:Hams-LVC-loss}
    \hat{H} = \hat{H}_\text{LVC} + \hat{H}_\text{loss},
\end{align}
where the LVC Hamiltonian $\hat{H}_\text{LVC}$ is given in Eq.~\ref{eq:Hams-LVC}, and the loss Hamiltonian is expressed as~\cite{Dou_2022, Arkajit_2022, Ying2023, Ying2023NanoP, Ying2024}
\begin{align} \label{eq:H-loss}
    \hat{H}_\text{loss} = \sum_j \frac{1}{2} \Big[\hat{P}^2_j + \Omega^2_j \Big(\hat{X}_j - \frac{C_j}{\Omega^2_j} (\hat{a} + \hat{a}^\dagger) \Big)^2 \Big], 
\end{align}
where $\hat{X}_{j}$ ($\hat{P}_{j}$) is the coordinate (momentum) operator for $j_{\mathrm{th}}$ far-field mode, with mode frequency $\Omega_j$ and coupling strength $C_j$ to the cavity mode. Note that the Hamiltonian in the form of Eq.~\ref{eq:H-loss} and its second-quantized form have also been widely used in the study of vibrational relaxation~\cite{Nitzan_MolPhys1973, Nitzan_JCP1973}. 
Here, we assume wide band approximation to the far-field mode frequencies which leads to a short correlation time, validating Markovian treatment of the cavity loss. 
To be specific, we assume the far-field modes as well as their coupling to the cavity mode can be described by a strictly Markovian Ohmic spectral density, 
\begin{align}
    J_{\mathrm{loss}}(\tilde{\omega}) = \frac{\pi}{2} \sum_j \frac{C^2_j}{\Omega_j} \delta(\tilde{\omega} - \Omega_j) = \Gamma \tilde{\omega} \exp(- \tilde{\omega}/\omega_\mathrm{m}),
\end{align}
with $\omega_\mathrm{m} \rightarrow \infty$, and $\Gamma = 1/\tau_\text{c}$ is the cavity loss rate. 

Following the approach developed by Leggett~\cite{Leggett_1984} and Garg, {\it et al}.~\cite{Ambegaokar_1985}, by performing a normal mode transformation, the Hamiltonian in Eq.~\ref{eq:Hams-LVC-loss} can be strictly mapped to an effective Hamiltonian expressed as follows~\cite{Ying2023}
\begin{widetext}
\begin{align} 
    \hat{H} &= \left(E_D + \frac{|g'_D|^2}{\hbar \omega} \right) |\text{D}\rangle \langle \text{D}| + \Big(E_A + \frac{|g'_A|^2}{\hbar \omega} + \sum_j \frac{\lambda^2_j}{\hbar \nu_j} \Big) |\text{A}\rangle \langle \text{A}|  + \left[H_{DA} + \frac{(g'_D + g'_A)t'_{DA}}{\hbar \omega} \right] |\text{D}\rangle \langle \text{A}| \notag\\
    &~~~ + \left[H_{AD} + \frac{(g'_D + g'_A)t'_{AD}}{\hbar \omega} \right] |\text{A}\rangle \langle \text{D}| + \sum_j \hbar \nu_j \hat{b}^\dagger_j \hat{b}_j + \sum_k \hbar \omega_k \hat{a}^\dagger_k \hat{a}_k \notag\\
    &~~~ + \Big( g'_D |\text{D}\rangle \langle \text{D}| + g'_A |\text{A}\rangle \langle \text{A}| + t'_{DA} |\text{D}\rangle \langle \text{A}| + t'_{AD} |\text{A}\rangle \langle \text{D}| \Big) \otimes \sum_k c_k (\hat{a}_k + \hat{a}^\dagger_k) + |\text{A}\rangle \langle \text{A}| \otimes \sum_j \lambda_j (\hat{b}_j + \hat{b}^\dagger_j), \label{eq:Hams-LVC-loss-eff}
\end{align}
\end{widetext}
where $\hat{a}^\dagger_k$ ($\hat{a}_k$) is the creation (annihilation) operator of the $k_{\mathrm{th}}$ normal mode, with mode frequency $\omega_k$ and coupling strength $c_k$ to the electronic states. 
The normal modes $\sum_k \hbar \omega_k \hat{a}^\dagger_k \hat{a}_k$ as well as their coupling to the electronic states can described by an effective spectral density function as follows~\cite{Ying2023},
\begin{equation}\label{eq:jeff}
    J_{\mathrm{eff}}(\tilde{\omega}) = \frac{\pi}{\hbar} \sum_{k} c^2_k \delta(\tilde{\omega} - \omega_k) = \frac{2\omega \Gamma \tilde{\omega}}{(\tilde{\omega}^2 - \omega^2)^2 + \Gamma^2 \tilde{\omega}^2},
\end{equation}
which is of a Brownian oscillator form (centered at $\omega$ and broadened by $\Gamma$), with reorganization energy
\begin{align} \label{eq:reorg-eff}
    \Lambda_\text{eff} = \frac{1}{\pi} \int_0^\infty d\tilde{\omega}~ \frac{J_{\mathrm{eff}}(\tilde{\omega})}{\tilde{\omega}} = \sum_k \frac{c^2_k}{\hbar \omega_k} = \frac{1}{\hbar \omega}. 
\end{align}

Based on the Hamiltonian in Eq.~\ref{eq:Hams-LVC-loss-eff} and follow the same procedures as deriving Eq.~\ref{eq:FFCF-2}, one obtains an analytic expression for the correlation function of FGR rate theory as follows
\begin{align} \label{eq:FFCF-3}
    C_{ff}(t) &= [\tilde{h}(t) + \tilde{g}(t)] \cdot e^{\tilde{f}(t)},
\end{align}
where $\tilde{h}(t)$, $\tilde{g}(t)$ and $\tilde{f}(t)$ are analogous to Eq.~\ref{eq:hgf_t}, expressed as following
\begin{widetext}
\begin{subequations} \label{eq:hgf_t_loss}
\begin{align}
    \tilde{h}(t) &= \Big\{ H_{DA} +  t'_{DA} g'_{DA} \cdot \hbar \omega \sum_k \frac{c^2_k}{\hbar \omega_k} \Big[- \cos(\omega_k t) + i \sin(\omega_k t) \coth(\frac{\beta \hbar \omega_k}{2}) \Big] \Big\} \notag\\
    &~~~ \times \Big\{ H_{AD} + t'_{AD} g'_{DA} \cdot \hbar \omega \sum_k \frac{c^2_k}{\hbar \omega_k} \Big[- \cos(\omega_k t) + i \sin(\omega_k t) \coth(\frac{\beta \hbar \omega_k}{2}) \Big] \Big\}, \\
    \tilde{g}(t) &= t'_{DA} t'_{AD} \sum_k c^2_k \Big\{ \cos(\omega_k t) \coth(\frac{\beta \hbar \omega_k}{2}) - i \sin(\omega_k t)  \Big\}, \\
    \tilde{f}(t) &= - |g'_{DA}|^2 (\hbar \omega)^2 \sum_k \frac{c^2_k}{(\hbar \omega_k)^2} \Big\{ [1 - \cos(\omega_k t)] \coth(\frac{\beta \hbar \omega_k}{2}) + i \sin(\omega_k t) \Big\} \notag\\
    &~~~ - \sum_j \frac{\lambda^2_j}{(\hbar \nu_j)^2} \Big\{ [1 - \cos(\nu_j t)] \coth(\frac{\beta \hbar \nu_j}{2}) + i \sin(\nu_j t) \Big\}. 
\end{align}
\end{subequations}
\end{widetext}
Note that for finite number of normal modes, $\omega_k$ and $c_k$ can be sampled from the effective spectral density in Eq.~\ref{eq:jeff}, see details in Appendix~\ref{apdx:comp-3}. On the other hand, for quasi-continuous spectral density functions $J_\text{vib}(\omega)$ and $J_\text{eff}(\omega)$, one can rewrite the discrete summation with respect to $j$ and $k$ in terms of integration, so that Eq.~\ref{eq:hgf_t_loss} becomes
\begin{widetext}
\begin{subequations}
\begin{align}
    \tilde{h}(t) &= \Big\{ H_{DA} +  t'_{DA} g'_{DA} \cdot \hbar \omega \cdot \frac{1}{\pi} \int_0^\infty d\tilde{\omega}~ \frac{J_\text{eff}(\tilde{\omega})}{\tilde{\omega}} \Big[- \cos(\tilde{\omega} t) + i \sin(\tilde{\omega} t) \coth(\frac{\beta \hbar \tilde{\omega}}{2}) \Big] \Big\} \notag\\
    &~~~ \times \Big\{ H_{AD} + t'_{AD} g'_{DA} \cdot \hbar \omega \cdot \frac{1}{\pi} \int_0^\infty d\tilde{\omega}~ \frac{J_\text{eff}(\tilde{\omega})}{\tilde{\omega}} \Big[- \cos(\tilde{\omega} t) + i \sin(\tilde{\omega} t) \coth(\frac{\beta \hbar \tilde{\omega}}{2}) \Big] \Big\}, \\
    \tilde{g}(t) &= t'_{DA} t'_{AD} \cdot \frac{\hbar}{\pi} \int_0^\infty d\tilde{\omega}~ J_\text{eff}(\tilde{\omega}) \Big\{ \cos(\tilde{\omega} t) \coth(\frac{\beta \hbar \tilde{\omega}}{2}) - i \sin(\tilde{\omega} t)  \Big\}, \\
    \tilde{f}(t) &= - \frac{1}{\hbar \pi} \int_0^\infty d\tilde{\omega}~ \frac{J_\text{vib}(\tilde{\omega}) + |g'_{DA}|^2 (\hbar \omega)^2 J_\text{eff}(\tilde{\omega})}{\tilde{\omega}^2} \Big\{ [1 - \cos(\tilde{\omega} t)] \coth(\frac{\beta \hbar \tilde{\omega}}{2}) + i \sin(\tilde{\omega} t) \Big\}.
\end{align}
\end{subequations}
\end{widetext}
It is straightforward to see that Eq.~\ref{eq:FFCF-3} reduces back to Eq.~\ref{eq:FFCF-2} when there is only one normal mode -- with $\omega_k = \omega$ and $c_k = 1$, or equivalently, adopting $J_\text{eff}(\tilde{\omega}) = (\pi / \hbar) \delta(\tilde{\omega} - \omega)$. 
We also emphasize that Eq.~\ref{eq:FFCF-3} is not restricted to the case with Markovian cavity loss, but rather general to non-Markovian electromagnetic environments, {\it i.e.}, applicable to arbitrary effective spectral density function $J_{\mathrm{eff}}(\tilde{\omega})$, which is also in accordance with the macroscopic QED framework~\cite{Hsu_JCP2019, Hsu_JCP2022-1, Hsu_JCP2022-2, Hsu_JPCL2022, Svendsen_2024, Hsu_JPCL2025}. 

A useful form of the general result in Eqs.~\ref{eq:FFCF-3}-\ref{eq:hgf_t_loss} is obtained in the intermediate temperature case -- a regime where most cavity polariton experiments under electronic strong coupling are operated~\cite{Zeng2023JACS, Lee2024JACS, Hutchinson2012ACIE, Munkhbat2018SA, Ng2015AOM, Mony2021AFM}. Here, the temperature is assumed high for the bath phonon modes and low with respect to the cavity modes, {\it i.e.}, $\hbar \omega_k \gg k_\text{B} T \gg \hbar \nu_j$, as is done in the Marcus-Jortner theory. Based on the correlation function in Eq.~\ref{eq:FFCF-3}, one can generalize the Marcus-Jortner theory in Eq.~\ref{eq:MJ_rate-C} to cases with many modes coupling, expressed as follows,
\begin{widetext}
\begin{align} \label{eq:MJ_rate_loss}
    k_{\text{D}\to \text{A}} &= \sqrt{\frac{\pi \beta}{\hbar^2 E_R}} \cdot e^{- \sum_k |g'_k|^2} \sum_{m=0}^\infty \sum_{k_1, \cdots, k_m} \frac{\prod_{\alpha=1}^m |g'_{k_\alpha}|^2}{m!} \Bigg\{ H_{DA} H_{AD} \cdot \exp\Big(- \beta \frac{[- \Delta G_0 - E_R - \sum_{\alpha=1}^m \hbar \omega_{k_\alpha} ]^2}{4 E_R}\Big) \notag\\
    &~~~ + \sum_{k_\beta} \big[c^2_{k_\beta} t'_{DA} t'_{AD} - g'_{k_\beta} c_{k_\beta} (H_{DA} t'_{AD} + H_{AD} t'_{DA}) \big] \cdot \exp\Big(- \beta \frac{[- \Delta G_0 - E_R - \sum_{\alpha=1}^m \hbar \omega_{k_\alpha} - \hbar \omega_{k_\beta}]^2}{4 E_R}\Big) \notag\\
    &~~~ + t'_{DA} t'_{AD} \sum_{k_\beta} \sum_{k_\gamma} c_{k_\beta} c_{k_\gamma} g'_{k_\beta} g'_{k_\gamma} \cdot \exp\Big(- \beta \frac{[- \Delta G_0 - E_R - \sum_{\alpha=1}^m \hbar \omega_{k_\alpha} - \hbar \omega_{k_\beta} - \hbar \omega_{k_\gamma}]^2}{4 E_R}\Big) \Bigg\}, 
\end{align}
\end{widetext}
which we refer to as the generalized Marcus-Jortner (GMJ) theory. The detailed derivation for Eq.~\ref{eq:MJ_rate_loss} is provided in Supplementary Material, Section IV. 
Similar as Eq.~\ref{eq:MJ_rate_conv-C}, one can also develop a convolution form for Eq.~\ref{eq:MJ_rate_loss}. 

For the case of fast cavity modes \& slow electron tunneling, one takes $g'_{DA} = 0$, Eq.~\ref{eq:MJ_rate_loss} then reduces to
\begin{align} \label{eq:MJ_rate_loss_A}
    &k_{\text{D}\to \text{A}} = \sqrt{\frac{\pi \beta}{\hbar^2 E_R}} \Bigg\{ H_{DA} H_{AD} \cdot \exp\Big(- \beta \frac{[- \Delta G_0 - E_R]^2}{4 E_R}\Big) \notag\\
    &~~~ + \sum_{k} c^2_{k} t'_{DA} t'_{AD} \cdot \exp\Big(- \beta \frac{[- \Delta G_0 - E_R - \hbar \omega_{k}]^2}{4 E_R}\Big) \Bigg\},
\end{align}
where only the $m = 0$ term in Eq.~\ref{eq:MJ_rate_loss} survives. 

On the other hand, for the case of slow cavity modes \& fast electron tunneling, one takes $t'_{DA} = t'_{AD} = 0$, Eq.~\ref{eq:MJ_rate_loss} then reduces to its first line only,
\begin{align} \label{eq:MJ_rate_loss_B}
    &k_{\text{D}\to \text{A}} = \sqrt{\frac{\pi \beta}{\hbar^2 E_R}} \cdot H_{DA} H_{AD} \cdot e^{- \sum_k |g'_k|^2} \notag\\
    &\times \sum_{m=0}^\infty \sum_{k_1, \cdots, k_m} \frac{\prod_{\alpha=1}^m |g'_{k_\alpha}|^2}{m!} \notag\\
    &\times \exp\Big(- \beta \frac{[- \Delta G_0 - E_R - \sum_{\alpha=1}^m \hbar \omega_{k_\alpha} ]^2}{4 E_R}\Big). 
\end{align}
In practical numerical calculations, $m$ shall be truncated at a finite value.


\section{Numerical Results} \label{sec:numerical}
In this section, we examine the implications of the results obtained above. In particular, we show that there will be a resonance effect of cavity modification to ET dynamics when the cavity mode frequency satisfies certain relation with $- \Delta G_0 - E_R$. And we further reveal an interesting regime where ET can induce photon emission. Finally, for lossy cavities, we show the cavity quality factor dependence of the ET rates. 

\subsection{Model parameters and methods}
We follow Ref.~\citenum{Nitzan_2019} by choosing two sets of parameters in Tabel~\ref{tab:par}, namely Model A (corresponding to fast cavity modes \& slow electron tunneling) and Model B (corresponding to slow cavity modes \& fast electron tunneling), respectively. 

\begin{table}[htbp]
    \renewcommand{\arraystretch}{1.5}
    \caption{Major Model Parameters.} \vspace{0.1 cm}
    \begin{tabular*}{1.0\columnwidth}{c| @{\extracolsep{\fill}} c @{\extracolsep{\fill}} c @{\extracolsep{\fill}} c @{\extracolsep{\fill}} c @{\extracolsep{\fill}} c}
        \hline\hline
        Model & $H_{DA}$ & $\hbar \omega$ & $t'_{DA}$ & $g'_{DA}$ & $E_R$ \\ [0.5ex]
        \hline 
        A & 245 cm$^{-1}$ & 2 eV & 69 cm$^{-1}$ & 0 & 1 eV \\ [0.5ex]
        \hline 
        B & 30 cm$^{-1}$ & 0.2 eV & 0.5 cm$^{-1}$ & 0.5 & 0.2 eV \\ [0.5ex]
        \hline\hline
    \end{tabular*}
    \label{tab:par}
\end{table}

We perform discretization for continuous spectral densities and numerical fast Fourier transform (FFT) to evaluate the FGR rate expressions. 
Computational details can be found in Appendix~\ref{apdx:comp}. 

\subsection{Numerical results of the FGR, Marcus, and MJ rate expressions}
We first check the FGR and its associated approximations under high- and low-temperature limits of the nuclei based on the Models A and B in Table~\ref{tab:par}. For simplicity, we focus on lossless cavities. 

\subsubsection{The thermal activation / potential surface-crossing limit}
We first look at the ET rates under the high-temperature limit ($\beta \to 0$) for both the cavity mode and the bath phonons. Fig.~\ref{fig:1}a presents numerical results of the ET rate as a function of the donor-acceptor energy gap $- \Delta G_0$ for Model A, with both outside the cavity and inside the cavity cases. The temperature is set as $T = 3\times 10^5$ K such that $k_\text{B}T \gg \hbar \omega,\, \hbar \omega_\text{c}$. 
For outside the cavity cases, the FGR results (Eq.~\ref{eq:FGR-LRT}, black open circles) agree well with the Marcus rates (Eq.~\ref{eq:Marcus}, black solid lines) across all the parameter regime explored. 
For inside the cavity cases, the FGR results (Eq.~\ref{eq:FGR-LRT}, blue dots) agree well with the Marcus rates (Eq.~\ref{eq:Marcus-A}, red dashed lines) across all the parameter regime explored. 
One sees that the ET rate inside the cavity is enhanced by approximately 5 times compared to outside the cavity cases. This is because the cavity mode plays the role of a fluctuating bridge within this parameter regime, providing additional reaction channels. 
Fig.~\ref{fig:1}b shows similar plot as Fig.~\ref{fig:1}a, but uses Model B parameters with $T = 3\times 10^4$ K, such that the condition $k_\text{B}T \gg \hbar \omega,\, \hbar \omega_\text{c}$ also holds. 
For inside the cavity cases, the FGR results (Eq.~\ref{eq:FGR-LRT}, blue dots) agree well with the Marcus rates (Eq.~\ref{eq:Marcus-C}, red dashed line) across all the parameter regime explored. 
One sees that when $-\Delta G_0$ is small, the ET rate inside the cavity is slightly suppressed compared to outside the cavity cases, this is because the cavity mode plays the same role as classical nuclei and increases the total reorganization energy. 
Note that the small off-diagonal coupling $t'_{DA} = t'_{AD} = 0.5$ cm$^{-1}$ cannot be neglected for this case, whose effect scales as $\beta^{-1}$ when $\beta \to 0$ (see Eq.~\ref{eq:k_0_C}). One sees that the Marcus results using Eq.~\ref{eq:Marcus-B} (green dashed line, which assumes $t'_{DA} = t'_{AD} = 0$) deviate from the FGR results; and Eq.~\ref{eq:Marcus-C} is needed to reach quantitative agreement with the FGR results. 

\subsubsection{Quantum cavity mode correction to classical nuclei}
Next, we look at the moderate temperature regime by fixing $T =$ 300 K, with high-temperature limit for the phonon bath as $k_\text{B}T \approx 206$ cm$^{-1}$ $\gg \hbar \omega_\text{c} = 20$ cm$^{-1}$; meanwhile, $k_\text{B}T \ll \hbar \omega$ holds for both Models A and B. 
Recall the discussions in Section~\ref{sec:FGR}, outside the cavity, the FGR will be reduced to Marcus theory (Eq.~\ref{eq:Marcus}); while in the presence of cavity mode coupling, one applies low-temperature limit to the cavity mode and the FGR reduces to the MJ theory (Eqs.~\ref{eq:MJ_rate_A}, \ref{eq:MJ_rate_B}, and \ref{eq:MJ_rate-C}). 
We numerically examine the FGR rate expression in Eq.~\ref{eq:FGR-LRT} (for both outside and inside the cavity cases), the Marcus rate expression in Eq.~\ref{eq:Marcus} (outside the cavity), and the MJ rate expression in Eqs.~\ref{eq:MJ_rate_A} and \ref{eq:MJ_rate_B} for Models A and B, respectively (inside the cavity). 

\begin{figure*}[htbp]
    \centering
    \includegraphics[width=0.75\linewidth]{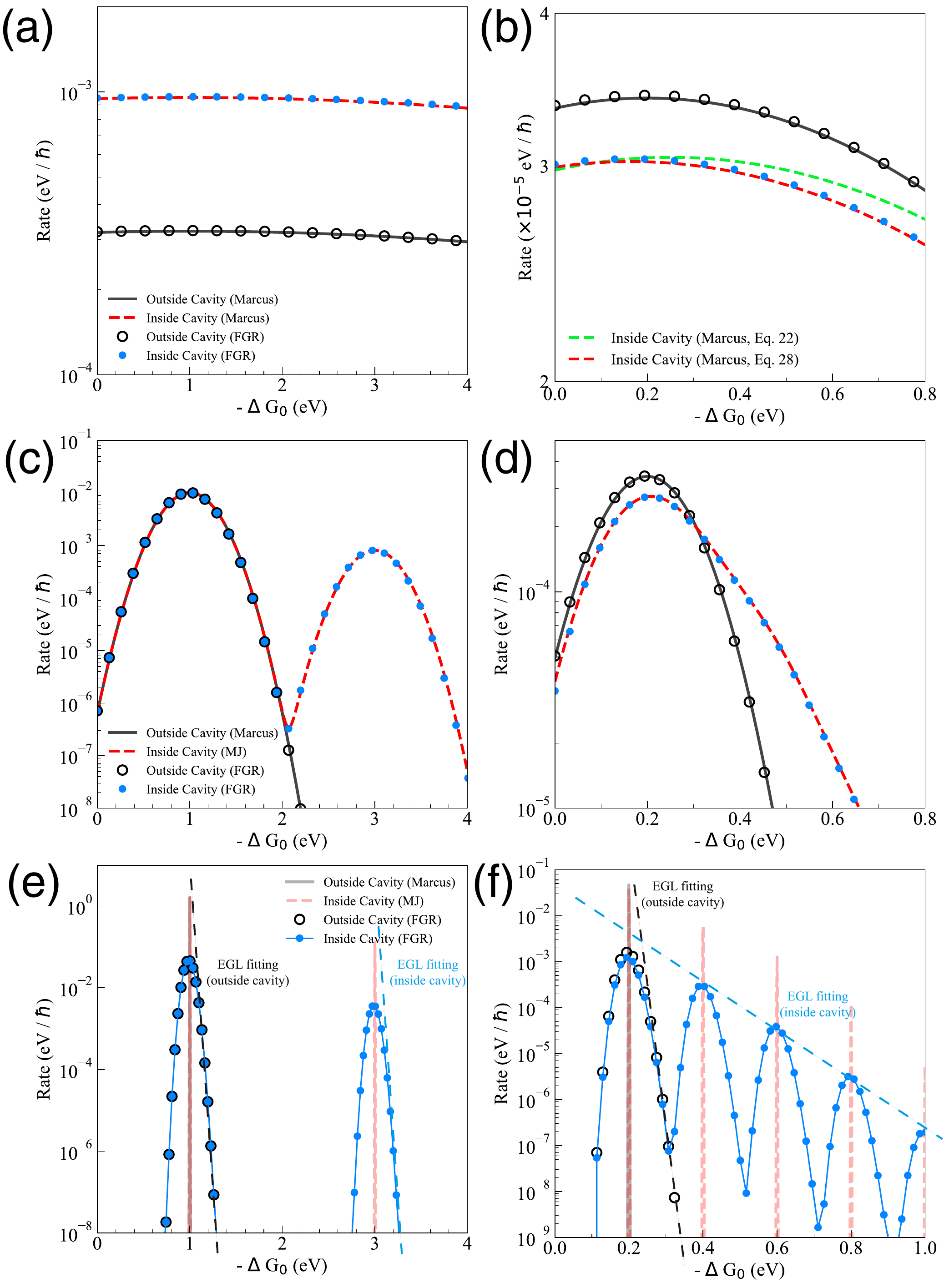}
    \caption{ET rates $k_{\text{D}\to \text{A}}$ obtained from FGR and various approximated rate expressions. (a) Model A under the high-temperature limit with $T = 3\times 10^5$ K. FGR results using Eq.~\ref{eq:FGR-LRT} (black open circles for outside cavity, and cyan dots for inside the cavity) are compared to Marcus results both outside cavity (Eq.~\ref{eq:Marcus}, gray solid lines) and inside the cavity (Eq.~\ref{eq:Marcus-A}, red dashed line) cases. (b) Model B under the high-temperature limit with $T = 3\times 10^4$ K. FGR results using Eq.~\ref{eq:FGR-LRT} are compared to Marcus results both outside cavity (Eq.~\ref{eq:Marcus}, gray solid lines) and inside the cavity (Eq.~\ref{eq:Marcus-B} in green dashed line, and Eq.~\ref{eq:Marcus-C} in red dashed line) cases. 
    (c) Model A under moderate temperature with $T = 300$ K. FGR results using Eq.~\ref{eq:FGR-LRT} are compared to Marcus results outside cavity (Eq.~\ref{eq:Marcus}, gray solid lines) and MJ results inside the cavity (Eq.~\ref{eq:Marcus-A}, red dashed line). (d) Model B under moderate temperature with $T = 300$ K. FGR results using Eq.~\ref{eq:FGR-LRT} are compared to Marcus results outside cavity (Eq.~\ref{eq:Marcus}, gray solid lines) and MJ results inside the cavity (Eq.~\ref{eq:Marcus-B}, red dashed line). 
    (e) Model A under low temperature with $T = 0$ K. FGR results using Eq.~\ref{eq:FGR-LRT} are compared to Marcus results outside cavity (Eq.~\ref{eq:Marcus}, gray solid lines) and MJ results inside the cavity (Eq.~\ref{eq:Marcus-A}, red dashed line). (f) Model B under low temperature with $T = 0$ K. FGR results using Eq.~\ref{eq:FGR-LRT} are compared to Marcus results outside cavity (Eq.~\ref{eq:Marcus}, gray solid lines) and MJ results inside the cavity (Eq.~\ref{eq:Marcus-B}, red dashed line).
    Note that in panels (e) and (f), $T = 0.01$ K has been applied to the Marcus / MJ rates in order to avoid singularities, and the EGL linear fitting results are shown in thin dashed lines (black for outside cavity, and cyan for inside cavity). }
    \label{fig:1}
\end{figure*}

Fig.~\ref{fig:1}c shows numerical results of the ET rate as a function of donor-acceptor energy gap $-\Delta G_0$ for both outside the cavity and inside the cavity cases based on Model A (corresponding to Fig.~3 of Ref.~\citenum{Nitzan_2019}).  
For outside the cavity cases, the FGR results (Eq.~\ref{eq:FGR-LRT}, black open circles) agree well with the Marcus rates (Eq.~\ref{eq:Marcus}, black solid lines) across all the parameter regime explored. 
For inside the cavity cases, the FGR results (Eq.~\ref{eq:FGR-LRT}, blue dots) agree well with the MJ rates (Eq.~\ref{eq:MJ_rate_A}, red dashed lines) across all the parameter regime explored. In particular, the cavity coupling results in an additional resonant peak centered at $-\Delta G_0 = E_R + \hbar\omega$, which can be well understood from Eq.~\ref{eq:MJ_rate_A}. 
Fig.~\ref{fig:1}d shows similar observations as Fig.~\ref{fig:1}c but uses Model B parameters (corresponding to Fig.~4 of Ref.~\citenum{Nitzan_2019}), and the MJ rates are obtained using Eq.~\ref{eq:MJ_rate_B}. 
As such, under the high-temperature limit for the phonon bath, the Marcus / MJ rate expressions serve as good approximations of the FGR. 

\subsubsection{The weak coupling limit and the energy gap law}
We further look at the low-temperature limit for both the phonon bath and the cavity mode, where the Marcus / MJ theory breaks down. In particular, we compare the EGL scaling relations between outside the cavity and inside the cavity cases using Models A and B at $T = 0$. 
Recall the discussions in Section~\ref{sec:fast-cavity} and Section~\ref{sec:slow-cavity}, Model A and Model B will exhibit distinct EGL scalings. 

Fig.~\ref{fig:1}e presents numerical results of the ET rate as a function of donor-acceptor energy gap $-\Delta G_0$ for both outside the cavity and inside the cavity cases based on Model A. 
For outside the cavity cases, the FGR results (Eq.~\ref{eq:FGR-LRT}, black open circles) significantly differ from the Marcus rates (Eq.~\ref{eq:Marcus}, silver solid lines). 
For inside the cavity cases, the FGR results (Eq.~\ref{eq:FGR-LRT}, blue dotted lines) also significantly differ from the MJ rates (Eq.~\ref{eq:MJ_rate_A}, red dashed lines). Note that $T = 0.01$ K has been applied to numerically implement the Marcus / MJ rate expressions in order to avoid singular values (as they reduces to Dirac $\delta$-functions under $T \to 0$). 
One clearly sees the break down of the Marcus / MJ theory under the low temperature limit. 
Meanwhile, based on the FGR rates, one can extract the EGL scaling relations via a linear fitting of the $k_\mathrm{D \to A}$ v.s. $- \Delta G_0$ plot as donor-acceptor energy gap increases. One sees from Fig.~\ref{fig:1}e that for Model A, the EGL scaling inside the cavity keeps the same as outside the cavity (manifested by the same slope), but shifted to the right-hand side by an amount of $\hbar \omega$, in accordance with the theoretical prediction in Eqs.~\ref{eq:EGL_A_1}-\ref{eq:EGL_A_2}. 

On the other hand, Fig.~\ref{fig:1}f shows similar plots as Fig.~\ref{fig:1}e but uses Model B parameters, and the MJ rates are obtained using Eq.~\ref{eq:MJ_rate_B}. Again, one sees that the Marcus / MJ rate expressions completely breaks down as $T \to 0$. 
In particular, inside the cavity, there are multiple sequential resonant peaks appearing at $- \Delta G_0 = E_R + m\hbar\omega$, where $m = 1, 2, \cdots$, which can be well understood from Eq.~\ref{eq:MJ_rate_B}. 
Moreover, one sees that the rate profiles outside / inside the cavity have different EGL scaling relation (manifested by different slopes under linear fitting in Fig.~\ref{fig:1}f), in accordance with the theoretical prediction in Eq.~\ref{eq:EGL_B_1}. 
Interestingly, the EGL scaling relation inside the cavity is also effectively captured by the MJ rate expression in Eq.~\ref{eq:MJ_rate_B} (red dashed lines) -- although it fails to give rise to the correct peak intensity and width. This is because the MJ theory also treats quantum mechanically the cavity mode DOF (which dominates the EGL for Model B under $T \to 0$). 

We note that the EGL scaling relations are also sensitive to the bath phonon characteristic frequency $\omega_\text{c}$. Details are presented in Appendix~\ref{apdx:wc-dep-EGL}. 

\subsection{Resonance effect}
Next, we investigate the resonance effect of cavity modification to ET dynamics, that is, with given $- \Delta G_0$ and $E_R$, there will be one (or multiple) specific cavity mode frequency $\omega$ that gives rise to maximal ET rate. 
For simplicity, we focus on $T =$ 300 K, assuming the high-temperature limit for the bath phonon modes where the Marcus / MJ rate expressions work well. And again we focus on the two Models A and B, but with rescaled light-matter coupling strength under a varying $\omega$ -- since the light-matter coupling strength will also depend on $\omega$ (see Eqs.~\ref{eq:LMCS} and \ref{eq:gDA}). We further assume the cavity mode volume $\Omega$ is fixed, so that $t'_{DA} \propto \sqrt{\omega}$ and $g'_{DA} \propto 1 / \sqrt{\omega}$.  
Since the cavity mode frequency $\omega$ is positive definite, we expect the resonance effect to appear only in the Marcus inverted regime (with $- \Delta G_0 - E_R > 0$). 

\begin{figure}[htbp]
    \centering
    \includegraphics[width=1.0\linewidth]{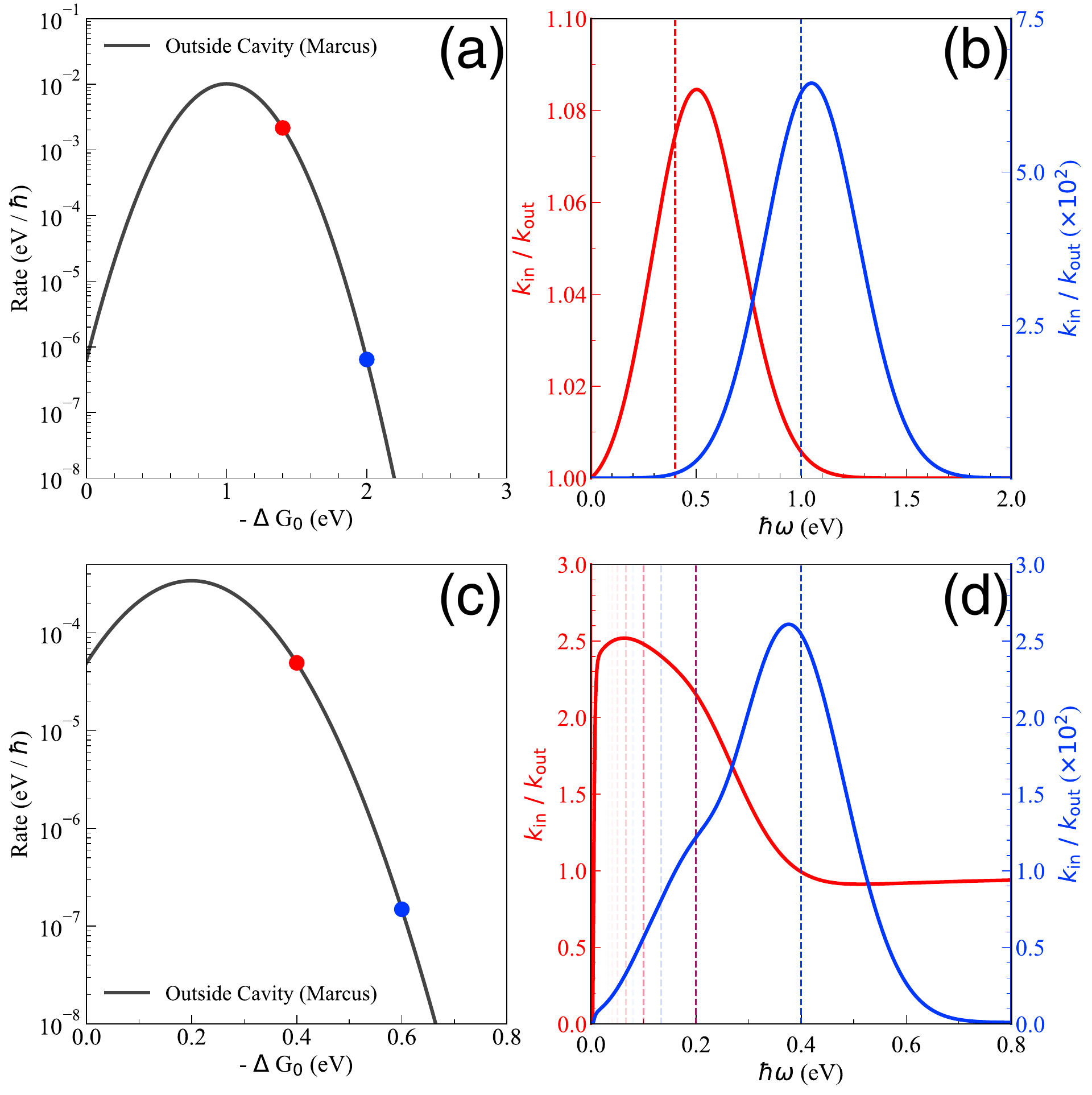}
    \caption{Resonance effect of the cavity modified ET rates. We fix $T =$ 300 K. 
    Panels (a)-(b) use the parameters of Model A while changing the cavity frequency $\omega$ and light-matter coupling strength $t'_{DA}$. Panel (a) shows the outside cavity ET rates ($k_\text{out}$) obtained using Marcus theory in Eq.~\ref{eq:Marcus}, from which we pick up two specific $-\Delta G_0$ values in the Marcus inverted regime, 1.4 eV (red) and 2.0 eV (blue), to explore the resonance effect, where the cavity modified ET rate $k_\text{in}$ is obtained using the MJ rate expression in Eq.~\ref{eq:MJ_rate_A}. Panel (b) shows the cavity rate modification $k_\text{in} / k_\text{out}$ by varying $\hbar \omega$. 
    Similarly, panels (c)-(d) use the parameters of Model B while changing $\omega$ and $g'_{DA}$. We pick up $-\Delta G_0 =$ 0.4 eV (red) and 0.6 eV (blue) and their corresponding $k_\text{out}$ in panel (c) to explore the resonance effect, where the cavity modified ET rate $k_\text{in}$ is obtained using the MJ rate expression in Eq.~\ref{eq:MJ_rate_B}. The corresponding cavity rate modification $k_\text{in} / k_\text{out}$ is shown in panel (d). The predicted resonance frequency is indicated by the red and blue dashed line(s). }
    \label{fig:8}
\end{figure}

For Model A, according to the MJ rate expression in Eq.~\ref{eq:MJ_rate_A}, one expects to see a maximal ET rate around $\omega = - \Delta G_0 - E_R$. 
To examine this prediction, we choose two specific $-\Delta G_0$ values, 1.4 eV and 2.0 eV, in the Marcus inverted regime, as is labeled in Fig.~\ref{fig:8}a (red and blue, respectively). Their corresponding (outside the cavity) ET rates can be read from Fig.~\ref{fig:8}a and is referred to as $k_\text{out}$. Then we include the cavity mode DOF to obtain the modified ET rate (referred to as $k_\text{in}$) using Eq.~\ref{eq:MJ_rate_A} -- note that according to Eq.~\ref{eq:LMCS}, the light-matter coupling strength is rescaled as $t'_{DA} = 69~\text{cm}^{-1} \cdot \sqrt{\hbar \omega / 2.0~\text{eV}}$. By varying the cavity mode frequency $\omega$, we plot the ratio of ET rate modification $k_\text{out} / k_\text{in}$ as a function of $\omega$ in Fig.~\ref{fig:8}b. 
For $-\Delta G_0 - E_R = 0.4$ eV (red curve), one sees a single peak with a maximal enhancement ratio $k_\text{out} / k_\text{in} \approx$ 1.084 around $\omega = 0.5$ eV. 
For $-\Delta G_0 - E_R = 1.0$ eV (blue curve), one sees a single peak with a maximal enhancement ratio $k_\text{out} / k_\text{in} \approx 6.5 \times 10^2$ around $\omega = 1.1$ eV. Note that the orders of magnitude huge enhancement observed here is reasonable due to a very small $k_\text{out}$ (see the blue dot in Fig.~\ref{fig:8}a), as is also revealed by Wei and Hsu~\cite{Hsu_JPCL2022}. 
Furthermore, the expected resonance conditions $\hbar \omega = - \Delta G_0 - E_R$ are also shown in Fig.~\ref{fig:8}b, with numerical values $\hbar \omega =$ 0.4 eV and 1.0 eV for the red and the blue dashed lines, respectively. One sees that both the $k_\text{out} / k_\text{in}$ peaks are blue shifted relative to the dashed lines, due to $t'_{DA} \propto \sqrt{\omega}$. Under the large detuning limit of $\hbar \omega \gg - \Delta G_0 - E_R$, the exponential term $\exp[- \beta (- \Delta G_0 - E_R - \hbar \omega)^2 / (4 E_R) ] \to 0$ (see the second line of Eq.~\ref{eq:MJ_rate_A}), thus $k_\text{out} / k_\text{in} \to 1$, going back to outside the cavity case. 

\begin{figure*}[htbp]
    \centering
    \includegraphics[width=1.0\linewidth]{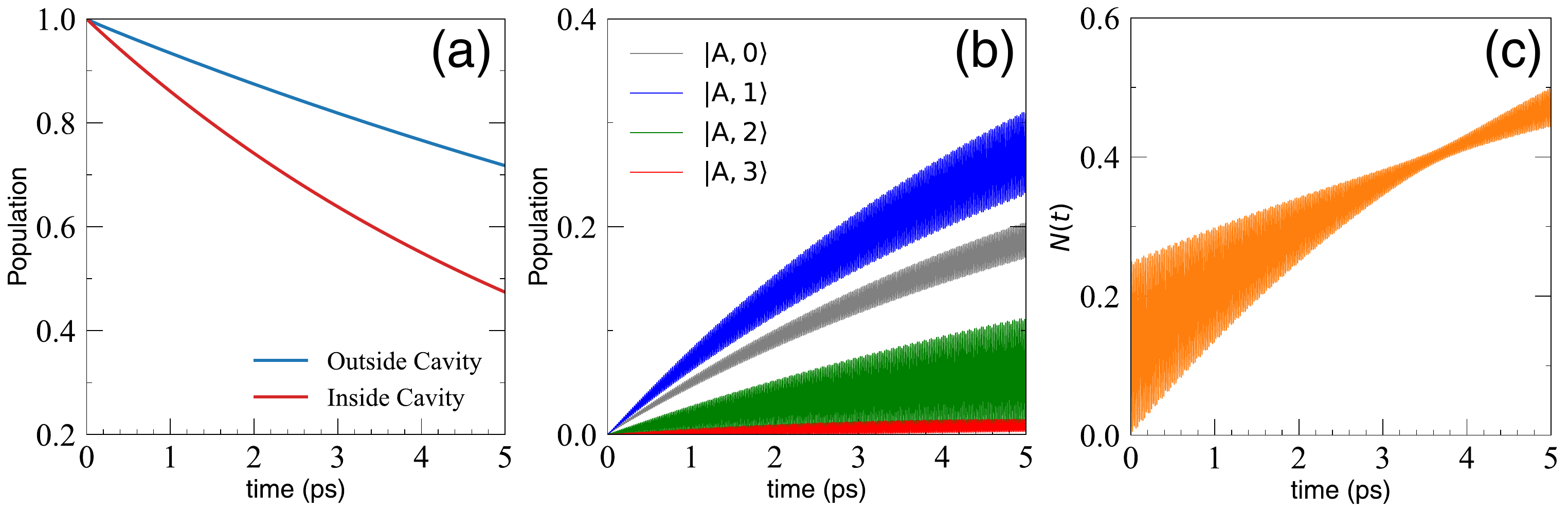}
    \caption{Quantum dynamics of the donor, acceptor states, as well as the photon number. Simulations are performed using the HEOM method. (a) Population dynamics of the donor state outside the cavity (blue) and inside the cavity (red). (b) Population dynamics of the photon-dressed acceptor states, $|\text{A}, 0 \rangle$ (gray), $|\text{A}, 1 \rangle$ (blue), $|\text{A}, 2 \rangle$ (green), and $|\text{A}, 3 \rangle$ (red). (c) Average photon number $N(t) = \langle \hat{a}^\dagger(t) \hat{a}(t) \rangle$ as a function of time. }
    \label{fig:3}
\end{figure*}

Fig.~\ref{fig:8}c-d shows similar plots as Fig.~\ref{fig:8}a-b, but uses Model B parameters and $k_\text{in}$ is obtained using Eq.~\ref{eq:MJ_rate_B} with rescaled light-matter coupling strength $g'_{DA} = 0.5 \times \sqrt{0.2~\text{eV} / (\hbar \omega)}$. Here, we choose two specific $-\Delta G_0$ values, 0.4 eV and 0.6 eV, in the Marcus inverted regime, as is labeled in Fig.~\ref{fig:8}c (red and blue, respectively). 
Fig.~\ref{fig:8}d shows more complicated resonance peaks -- as we expect the rate is maximized when $\omega = (- \Delta G_0 - E_R) / m$, $m = 1, 2, \cdots$ according to Eq.~\ref{eq:MJ_rate_B}. 
For example, for the red curve, the peak shape is a combination of multiple peaks at $0.2$ eV, $0.1$ eV, $0.0666$ eV, $0.05$ eV, $\cdots$ (for $m = 1, 2, 3, 4, \cdots$), as is indicated by the red dashed lines. And the curve also plateaus to $k_\text{out} / k_\text{in} \to 1$ when $\omega \gg - \Delta G_0 - E_R$ because $g'_{DA} \to 0$. 
On the other hand, the blue curve shows a clear shoulder at $\omega = 0.2$ eV alongside the major peak at $\omega = 0.4$ eV. 

\subsection{ET induced photon emission}
There is one intriguing aspect of the case of the slow cavity mode. In this case, the cavity mode plays a role similar to other molecular vibrations. It can be populated ({\it i.e.}, electronic energy converts to photon) during the transition (if $- \Delta G_0 > 0$) or excite it (using external source) may help the transition (if $- \Delta G_0 < 0$). However, unlike other slow modes whose excitation is thermal, exciting this mode may involve photons in the far field. For example, if the cavity mode is populated during the electronic transition, this may give rise to electron-transfer-induced photon emission. Here, we investigate this possibility using quantum dynamics simulations with the hierarchical equations of motion (HEOM) approach~\cite{Tanimura1990, YAYan_2004, Xu_2005, Yan_2014, Tanimura_2020}. 

To be specific, we focus on Model B and choose $-\Delta G_0 =$ 0.4 eV, $T =$ 300 K. To facilitate the HEOM simulations, we use the Drude-Lorentz cutoff function for the phonon bath spectral density (instead of the exponential cutoff form in Eq.~\ref{eq:Ohmic}), reading as
\begin{equation} \label{Drude}
    J_\text{vib}(\tilde{\omega}) = \frac{2 E_R\, \omega_\mathrm{c}\,  \tilde{\omega}}{\tilde{\omega}^2 + \omega_\mathrm{c}^2},
\end{equation}
where we choose the phonon characteristic frequency $\hbar \omega_\mathrm{c} = 20$ cm$^{-1}$, and $E_R = 0.2$ eV for Model B. Details on the HEOM method and its numerical implementations are provided in Supplementary Material, Section VI. To this end, one obtains the time-dependent reduced density matrix $\hat{\rho}_\text{S}(t)$ of the hybrid electron-photon subsystem. 

Fig.~\ref{fig:3} shows quantum dynamics of of the donor, acceptor states, as well as the photon numbers. In particular, Fig.~\ref{fig:3}a shows the donor population dynamics $P_D(t) = \mathrm{Tr}[|\text{D}\rangle \langle \text{D} | \hat{\rho}_\text{S}(t)]$. One sees that when coupling to the cavity mode, the donor population $P_D(t)$ (red solid line) decreases much faster than the cavity free case (blue solid line). 
Based on the population dynamics in Fig.~\ref{fig:3}a, one can extract the rate constant following the procedure in Supplementary Material, Section VI-E. 
The ET rate constant outside the cavity is $k_\text{out}^\text{HEOM} = 4.34 \times 10^{-5}$ eV$/ \hbar$, in good agreement with the corresponding Marcus theory result $k_\text{out}^\text{Marcus} = 4.93 \times 10^{-5}$ eV$/ \hbar$. 
On the other hand, the calculated ET rate constant inside the cavity as $k_\text{in}^\text{HEOM} = 0.985 \times 10^{-4}$ eV$/ \hbar$, in good agreement with the corresponding MJ theory result $k_\text{in}^\text{MJ} = 1.06 \times 10^{-4}$ eV$/ \hbar$. 

Fig.~\ref{fig:3}b shows the population dynamics of the first four photon-dressed acceptor states, $|\text{A}, 0 \rangle$ (gray), $|\text{A}, 1 \rangle$ (blue), $|\text{A}, 2 \rangle$ (green), and $|\text{A}, 3 \rangle$ (red), respectively. 
One sees that the population oscillates fast over time (due to the counter-rotating term is preserved), showing a steady increasing trend. In particular, the $|\text{A}, 1 \rangle$ state population (blue) increases fastest, while higher photon number states (green and red) are less prominent. 
The population dynamics of photon-dressed acceptor states clearly indicates photon generation during the ET process. 
Finally, Fig.~\ref{fig:3}c shows the averaged photon number dynamics, $N(t) = \langle \hat{a}^\dagger (t) \hat{a} (t) \rangle = \mathrm{Tr}[\hat{a}^\dagger \hat{a} \hat{\rho}_\text{S}(t)]$. The photon number also oscillates and increases over time, being a direct indicator of photon generation.  
Meanwhile, it is worth mentioning that this ET induced photon emission effect is also clearly reflected in Fig~\ref{fig:6}d (under $T =$ 0 K) -- one sees that as $-\Delta G_0$ increases, there will be multiple subsequent peaks in the cavity modified ET rate (blue curve), corresponding to resonant transitions from the zero-photon dressed donor state $|\text{D}, 0\rangle$ to $m$-photon dressed acceptor state $|\text{A}, m\rangle$, where $m = 1, 2, \cdots$ denotes the photon number. 

\subsection{Cavity quality factor dependence}
Introducing cavity loss, the ET rate can be evaluated using the FGR expression in Eq.~\ref{eq:FFCF-3}. 
For convenience, we take the same model A and B parameters in Table~\ref{tab:par} while introduce $\Gamma$ as the cavity loss rate. And we focus on $T =$ 300 K.
The cavity quality factor is defined as 
\begin{align}
    \mathcal{Q} = \omega / \Gamma. 
\end{align}
We report the ET rate changing as $\mathcal{Q}$ decreases (from $\infty$ to zero). 

\begin{figure*}[htbp]
    \centering
    \includegraphics[width=0.8\linewidth]{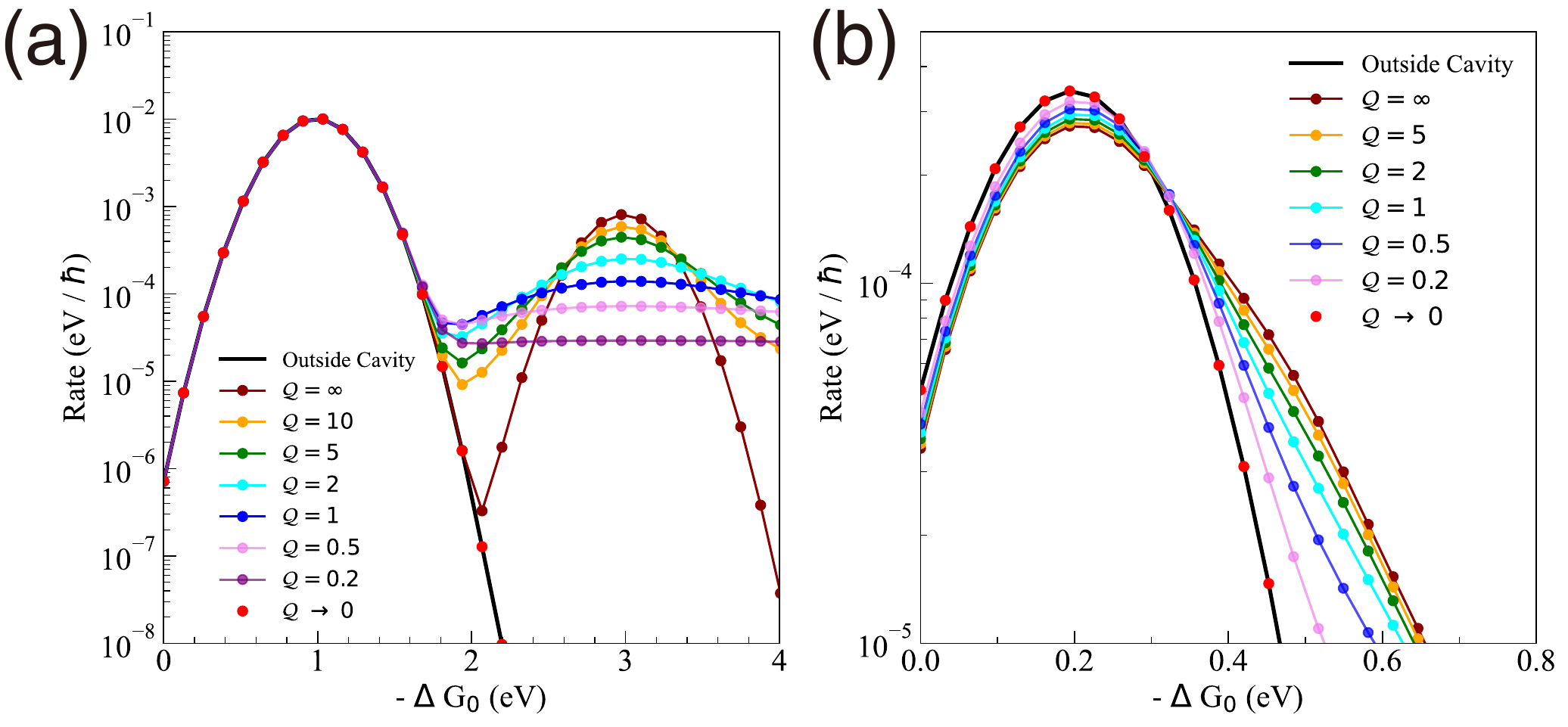}
    \caption{ET rate obtained from FGR using Eq.~\ref{eq:FFCF-3} inside lossy cavities with various $\mathcal{Q}$ factors. Here, we fix $T =$ 300 K. The parameters are taken from (a) Model A, and (b) Model B, respectively. }
    \label{fig:4}
\end{figure*}

Fig.~\ref{fig:4} shows the numerical result of FGR rates with loss (using Eq.~\ref{eq:FFCF-3}) under different $\mathcal{Q}$, where Fig.~\ref{fig:4}a and Fig.~\ref{fig:4}b correspond to Model A and B, respectively. 
One sees that the cavity modification effects are gradually weakened as $\mathcal{Q}$ decrease, showing an asymptotic trend to the outside the cavity rate (black solid curve). 
In particular, the ET rate reduces to outside the cavity case as $\mathcal{Q} \to 0$ (red dots). Here, numerically, $\mathcal{Q} = 2 \times 10^{-6}$ is taken to approach the $\mathcal{Q} \to 0$ limit. 
Under the high-temperature limit for the photon bath ($T =$ 300 K), we also tested the GMJ rate expressions in Eqs.~\ref{eq:MJ_rate_loss_A} and \ref{eq:MJ_rate_loss_B} for Models A and B, respectively, which show excellent agreement with the FGR results across all parameter regimes explored. Details are presented in Appendix~\ref{apdx:MJ_Rate_loss}. 

\section{Discussions} \label{sec:conclusion}
To summarize, in this work, we have developed a unified theoretical framework for cavity-modified electron transfer based on Fermi's golden rule rate theory. Starting from the polaron-transformed Hamiltonian, we derived analytic expressions for the force–force correlation function, enabling rate theories that remain valid across temperature regimes and cavity mode time scales. The resulting expressions naturally reproduce the Marcus and Marcus–Jortner formulas under the appropriate limits, while in the low-temperature regime they capture the emergence of the energy gap law. By extending the theory to lossy cavities through an effective Brownian oscillator spectral density, we established closed-form results that account for finite photon lifetimes / cavity quality factors. Numerical analyses highlighted two key consequences of cavity coupling: the resonance enhancement of ET when the cavity frequency matches relevant energetic parameters, and the possibility of electron-transfer-induced photon emission in the slow-cavity regime. Together, these results demonstrate how confined electromagnetic fields reshape charge-transfer dynamics, and provide a general foundation for exploring new strategies to control molecular reactivity in nanophotonic environments. 

Although all the numerical demonstrations in this work are based on Models A and B in Tabel~\ref{tab:par}, corresponding to two extremes of fast cavity mode \& slow electron tunneling and slow cavity mode \& fast electron tunneling, respectively; we emphasize that the FGR rate expression in Eq.~\ref{eq:FGR-LRT} (with the correlation function in Eq.~\ref{eq:FFCF-2} or the lossy version in Eq.~\ref{eq:FFCF-3}) and its high-temperature approximations -- the Marcus or Marcus-Jortner rate expression in Eq.~\ref{eq:Marcus-C}, \ref{eq:MJ_rate-C} or Eq.~\ref{eq:MJ_rate_loss}, are generally valid (under the nonadiabatic limit) in the intermediate regime where the cavity mode and electron tunneling have similar time scale. In these cases, the cross terms of $t'_{DA}$ and $g'_{DA}$ in these expressions should be important and cannot be dropped (as is already shown in Fig.~\ref{fig:1}b). 
We also emphasize that the theoretical framework is in principle not restricted to ET rates modified by the confined electromagnetic fields, but rather general to arbitrary environments in the scope of non-Condon ET. In particular, the FGR with cavity loss (or the approximated GMT expression in Eq.~\ref{eq:MJ_rate_loss}) developed in this work can be directly applied to study cavity modified ET rates beyond the single mode approximation, {\it i.e.}, cases with given dispersion relation of the cavity photonic modes, and being compatible with the macroscopic QED framework~\cite{Hsu_JCP2019, Hsu_JCP2022-1, Hsu_JCP2022-2, Hsu_JPCL2022, Svendsen_2024, Hsu_JPCL2025} for realistic electromagnetic environments. 
On the other hand, based on the analytic correlation functions derived in this paper, one can straightforwardly work out the non-Markovian generalization of the equilibrium rate constants using non-equilibrium Fermi's golden rule~\cite{Sun_Geva_2016, Sun_Geva_2016_2} (NE-FGR), or construct the perturbative quantum master equations. 
It would also be interesting to apply the current theory to study cavity modified charge transport in extended systems~\cite{Pupillo_PRL2017, Scholes_JCP2022}. 

The theoretical analysis presented in this paper adds to and generalizes the ongoing theoretical discussions of the possible effects of optical cavity environment on molecular electron transfer. Unfortunately, so far no direct experimental observation pertaining to these predictions were made. 
Below, we outline several possible signatures of these effects that might be amenable to experimental observations. 
(1) The resonance effect. In the optical domain, one could embed a molecular donor–acceptor pair with a well-defined charge-transfer step inside a tunable microcavity and extract the ET kinetics (for example, from transient absorption or time-resolved photoluminescence) while mechanically scanning the cavity resonance. A non-monotonic dependence of the extracted ET rate constant on cavity frequency would directly indicate the predicted resonance. (2) ET induced photon emission. This could be probed either optically, by monitoring spectrally resolved cavity leakage following photo excitation of the donor, or electrically, by integrating a biased donor–acceptor junction into a high-$\mathcal{Q}$ confined electromagnetic mode and correlating the measured current with photon counts at the cavity frequency. Observation of such photon–current correlations would constitute direct evidence of electron-transfer-driven light emission.

Despite the merits of the FGR rate theory (that covers all temperature regimes, fast and slow cavity modes, lossy and lossless cavities), there are several limitations in the current work and requires future efforts to address them. To be specific, 
\begin{itemize}
    \item {\bf Strong Coupling Scenario}. The current (FGR, Marcus or Marcus-Jortner) theory assumes ET to occur in the nonadiabatic limit where the nuclei can be treated with harmonic free energy surfaces, requiring a relatively small electron tunneling parameter (or more rigorously, a small Landau-Zener nonadiabaticity parameter~\cite{Nitzan, Manolopoulos_2019, Rabani_2023}). 
    However, as the adiabaticity increases and enters into the adiabatic limit ({\it e.g.} fast cavity mode under the ESC regime), the FGR rate theory gradually breaks down and one expects to use the transition state theory theory with the Born–Oppenheimer surfaces instead~\cite{Gladkikh_2005, Rabani_2023}. 
    As such, if one explores the effect of light-matter coupling strength by gradually increasing $t'_{DA}$ ({\it e.g.}, for Model A) that causes nonadiabatic to adiabatic crossover, attention to the adiabaticity needs to be paid. 
    On the other hand, changing $g'_{DA}$ for Model B remains trivial as it only influences the total reorganization energy rather than the adiabaticity. 
    \item {\bf Collective Effect}. The current theory and simulation assume that a single donor-acceptor pair is coupled to the cavity, while the experiments are usually operated under a collective coupling regime where a large ensemble of molecules coupled to a cavity mode. 
    It remains to be an open question that will 2 (or $N$) molecules behave differently from one molecule~\cite{Herrera_Spano_PRL, Mauro_PRB2021, Wellnitz_JCP2021, Chen_JCP2024}. Still, collectivity, if it exists, is the most interesting aspect of the whole phenomenon, and the only observable whose collective behavior is well understood is the Rabi splitting. 
    \item {\bf Ab Initio Modeling}. The current theory and simulations are based on model systems without atomistic details. 
    Future efforts shall be focusing on applying the theory to realistic reaction systems at an ab initio level. For example, performing electronic structure and molecular dynamics simulations to obtain the electron tunneling and phonon spectral density, respectively; and to obtain mode distribution for realistic electromagnetic environments via the macroscopic QED approach.  
\end{itemize}

\section*{Acknowledgements}
This work was supported by the European Research Council under ERC-2024-SyG-101167294; UnMySt. 
W.Y. thanks valuable discussions with Eitan Geva, Yifan Lai, and Pengfei Huo upon NE-FGR, and with Nadine Bradbury upon numerical implementation of FFT. We thank Dvira Segal for valuable discussions. 

\section*{Data Availability}
\noindent The data that support the findings of this work are available in \url{https://github.com/Okita0512/Cavity-ET-single}.

\appendix

\section{Derivation of the Hamiltonian in Eq.~\ref{eq:Hams-LVC}} \label{apdx:Hams-original}
The Hamiltonian of ET inside a cavity (with a single cavity mode) reads as (c.f. Eq. 16 of Ref.~\citenum{Nitzan_2019} with original notations, but combine same type of terms)
\begin{widetext}
\begin{align} \label{eq:Hams-original}
    \hat{H} &= (E_D + \hbar \omega |g_D|^2) |\text{D}\rangle \langle \text{D}| + (E_A + \hbar \omega |g_A|^2 + \sum_j \frac{\lambda^2_j}{\hbar \nu_j}) |\text{A}\rangle \langle \text{A}| + \hbar \omega |t_{DA}|^2 \hat{I} + [H_{DA} - \hbar \omega(g_D + g_A)t_{DA}] |\text{D}\rangle \langle \text{A}| \notag\\
    &~~~ + [H_{AD} - \hbar \omega(g_D + g_A)t_{AD}] |\text{A}\rangle \langle \text{D}| + \sum_j \hbar \nu_j \hat{b}^\dagger_j \hat{b}_j + \hbar \omega \hat{a}^\dagger \hat{a} \\
    &~~~+ \hbar \omega g_D (\hat{a} - \hat{a}^\dagger) \otimes |\text{D}\rangle \langle \text{D}| + [\hbar \omega g_A (\hat{a} - \hat{a}^\dagger) +  \sum_j \lambda_j (\hat{b}_j + \hat{b}^\dagger_j)] \otimes |\text{A}\rangle \langle \text{A}| + \hbar \omega (\hat{a} - \hat{a}^\dagger) \otimes (t_{DA}|\text{D}\rangle \langle \text{A}| + t_{AD} |\text{A}\rangle \langle \text{D}|), \notag
\end{align}
\end{widetext}
which is obtained by performing PZW gauge transformation to the minimal coupling Hamiltonian under the Coulomb gauge. 
In Eq.~\ref{eq:Hams-original}, $\{g_D, g_A\}$ and $\{t_{DA}, t_{AD}\}$ are the diagonal and off-diagonal light-matter coupling parameters, respectively, see details in Eq.~17 of Ref.~\citenum{Nitzan_2019}. Other symbols keep in consistence with Eq.~\ref{eq:Hams-LVC}. 
Note that the phonon bath reorganization energy term $\sum_j \frac{\lambda^2_j}{\hbar \nu_j} |\text{A}\rangle \langle \text{A}|$ has been added to Eq.~\ref{eq:Hams-original} to preserve translational invariance of the Hamiltonian. 

In order to give rise to the Pauli-Fierz Hamiltonian in Eq.~\ref{eq:Hams-LVC}, we further apply a unitary transform of phase shift to the Hamiltonian in Eq.~\ref{eq:Hams-original}, $\hat{U}_\phi = \exp(-i \phi \hat{a}^\dagger \hat{a})$, such that $\hat{U}_\phi \hat{a} \hat{U}^\dagger_\phi = e^{i\phi} \hat{a}$, and $\hat{U}_\phi \hat{a}^\dagger \hat{U}^\dagger_\phi = e^{- i\phi} \hat{a}^\dagger$. Here we choose $\phi = - \pi / 2$ for convenience, such that $\hat{a} \to - i\hat{a}$, and $\hat{a}^\dagger \to i \hat{a}^\dagger$. 
For convenience, we further redefine the coupling parameters by absorbing the imaginary unit as well as $\hbar \omega$ as follows (c.f. Eq.~17 of Ref.~\citenum{Nitzan_2019})
\begin{subequations} \label{eq:LMCS}
\begin{align}
    g_D \to g'_D &= -i \hbar \omega g_D = \sqrt{\frac{\hbar \omega }{2\Omega \epsilon_0}} \vec{d}_{DD} \cdot \vec{\xi}, \\
    g_A \to g'_A &= -i \hbar \omega g_A = \sqrt{\frac{\hbar \omega }{2 \Omega \epsilon_0}} \vec{d}_{AA} \cdot \vec{\xi}, \\
    t_{DA} \to t'_{DA} &= - i \hbar \omega t_{DA} = \sqrt{\frac{\hbar \omega}{2\Omega \epsilon_0}} \vec{d}_{DA} \cdot \vec{\xi}, \\
    t_{AD} \to t'_{AD} &= - i \hbar \omega t_{AD} = \sqrt{\frac{\hbar \omega}{2 \Omega \epsilon_0}} \vec{d}_{AD} \cdot \vec{\xi},
\end{align}
\end{subequations}
where $\Omega$ represents the cavity mode volume, $\epsilon_0$ is the permittivity inside the cavity, $\vec{d}_{DD}$ ($\vec{d}_{AA}$) is the permanent dipole moment associated to the donor (acceptor) states, which is essentially the donor (acceptor) position, $\vec{d}_{DA} = \vec{d}_{AD}^*$ is the transition dipole moment between the donor and acceptor orbitals, and $\vec{\xi}$ denotes the cavity field polarization vector. 
After performing the phase shift using $\hat{U}_\phi$, the Hamiltonian in Eq.~\ref{eq:Hams-LVC} is immediately obtained by dropping the $\frac{|t'_{DA}|^2}{\hbar \omega} \hat{I}$ term, as the constant identity operator does not influence the ET dynamics.

\section{An alternative form of the FGR rate expression} \label{apdx:FGR-2}
There are multiple ways to formulate the FGR rate. The most classic approach is given by the Fourier transform of the force-force correlation function (based on the PT Hamiltonian in Eq.~\ref{eq:Hams-PT-3}), as is discussed in Eq.~\ref{eq:FGR-LRT} of the main text. 
An alternative approach has been widely discussed by Geva, {\it et al.}~\cite{Sun_Geva_2016, Sun_Geva_2016_2, Geva_JCP2023, Geva_JPCL2022, Geva_JPCC2023}, where the Hamiltonian is separated into the diagonal and off-diagonal parts based on the LVC Hamiltonian in Eq.~\ref{eq:Hams-LVC},
\begin{align}
    \hat{H} = \hat{H}_{D} |\text{D}\rangle \langle \text{D}| + \hat{H}_{A} |\text{A}\rangle \langle \text{A}| + \hat{H}_{DA} |\text{D}\rangle \langle \text{A}| + \hat{H}_{AD} |\text{A}\rangle \langle \text{D}|, 
\end{align}
with the diagonal part
\begin{subequations}
\begin{align}
    \hat{H}_{D} &= E_D + \frac{|g'_D|^2}{\hbar \omega} + \hat{h}_\text{B} + g'_D (\hat{a} + \hat{a}^\dagger), \label{eq:HD-1}\\
    \hat{H}_{A} &= E_A + \frac{|g'_A|^2}{\hbar \omega} + \sum_j \frac{\lambda^2_j}{\hbar \omega_j} + \hat{h}_\text{B} + g'_A (\hat{a} + \hat{a}^\dagger) \notag\\
    &~~~ + \sum_j \lambda_j (\hat{b}_j + \hat{b}^\dagger_j),  \label{eq:HA-1}
\end{align}
\end{subequations}
and the off-diagonal part
\begin{subequations}
\begin{align}
    \hat{H}_{DA} &= H_{DA} + \frac{(g'_D + g'_A)t'_{DA}}{\hbar \omega} + t'_{DA} (\hat{a} + \hat{a}^\dagger), \\
    \hat{H}_{AD} &= H_{AD} + \frac{(g'_D + g'_A)t'_{AD}}{\hbar \omega} + t'_{AD} (\hat{a} + \hat{a}^\dagger), 
\end{align}
\end{subequations}
respectively. Here, the bath Hamiltonian $\hat{h}_\text{B} = \hbar \omega \hat{a}^\dagger \hat{a} + \sum_j \hbar \nu_j \hat{b}^\dagger_j \hat{b}_j$. 
The non-Condon equilibrium FGR theory~\cite{Sun_Geva_2016} states that
\begin{align} \label{eq:FGR-NCD}
    k_{\text{D}\to \text{A}} = \frac{1}{\hbar^2} \int_{-\infty}^{\infty} dt~ C_{DA}(t),
\end{align}
with a time-correlation function
\begin{align} \label{eq:CDA}
    C_{DA}(t) &= \text{Tr}\Big[\hat{\rho}^\text{eq}_D e^{i\hat{H}_D t / \hbar} \hat{H}_{DA} e^{- i\hat{H}_A t / \hbar} \hat{H}_{AD} \Big],
\end{align}
where $\hat{\rho}^\text{eq}_D = e^{-\beta\hat{H}_D} / \text{Tr}[e^{-\beta\hat{H}_D}]$. 
We prove below that the FGR rate expression in Eq.~\ref{eq:FGR-NCD} is equivalent to Eq.~\ref{eq:FGR-LRT}. 

To begin with, one notices that (c.f. Eq.~\ref{eq:HD-1} and \ref{eq:HA-1})
\begin{subequations} \label{eq:HD-unitary}
\begin{align}
    \hat{H}_D &= E_D + e^{\hat{S}_D} \hat{h}_\text{B} e^{- \hat{S}_D}, \\
    \hat{H}_A &= E_A + e^{\hat{S}_A} \hat{h}_\text{B} e^{- \hat{S}_A}
\end{align}
\end{subequations}
where $\hat{S}_D := - \frac{g'_D}{\hbar \omega} (\hat{a}^\dagger - \hat{a})$ and $\hat{S}_A := - \frac{g'_A}{\hbar \omega} (\hat{a}^\dagger - \hat{a}) - \sum_j \frac{\lambda_j}{\hbar \nu_j} (\hat{b}^\dagger_j - \hat{b}_j)$ are the shift operators. The proof is provided in Supplementary Material, Section VII. 
Furthermore,
\begin{subequations} \label{eq:UD-unitary}
\begin{align}
    e^{i\hat{H}_D t / \hbar} &= e^{i E_D t / \hbar} \cdot e^{\hat{S}_D} e^{i \hat{h}_\text{B}t / \hbar} e^{- \hat{S}_D}, \\
    e^{- i\hat{H}_A t / \hbar} &= e^{- i E_A t / \hbar} \cdot e^{\hat{S}_A} e^{- i \hat{h}_\text{B}t / \hbar} e^{- \hat{S}_A}, \\
    \hat{\rho}^\text{eq}_D &= e^{\hat{S}_D} \hat{\rho}^\text{eq}_\text{B} e^{- \hat{S}_D}. 
\end{align}
\end{subequations}
Using the results in Eqs.~\ref{eq:HD-unitary}-\ref{eq:UD-unitary}, one has
\begin{widetext}
\begin{align} \label{eq:CDA-Cff}
    C_{DA}(t) &= \text{Tr}\Big[\hat{\rho}^\text{eq}_D e^{i\hat{H}_D t / \hbar} \hat{H}_{DA} e^{- i\hat{H}_A t / \hbar} \hat{H}_{AD} \Big] \notag\\
    &= e^{i E_D t / \hbar} e^{- i E_A t / \hbar} \times \text{Tr}\Big[ e^{\hat{S}_D} \hat{\rho}^\text{eq}_\text{B} e^{- \hat{S}_D} \cdot e^{\hat{S}_D} e^{i \hat{h}_\text{B}t / \hbar} e^{- \hat{S}_D} \cdot \hat{H}_{DA} \cdot  e^{\hat{S}_A} e^{- i \hat{h}_\text{B}t / \hbar} e^{- \hat{S}_A} \cdot \hat{H}_{AD} \Big] \notag\\
    &= e^{i (E_D - E_A) t / \hbar} \times \text{Tr}\Big[ e^{i \hat{h}_\text{B}t / \hbar} \cdot e^{- \hat{S}_D} \hat{H}_{DA} e^{\hat{S}_A} \cdot e^{- i \hat{h}_\text{B}t / \hbar} \cdot e^{- \hat{S}_A} \hat{H}_{AD} e^{\hat{S}_D} \cdot \hat{\rho}^\text{eq}_\text{B} \Big] \notag\\
    &= e^{i (E_D - E_A) t / \hbar} \times \text{Tr} [e^{i\hat{h}_\text{B}t / \hbar} \hat{F}_\text{DA} e^{- i\hat{h}_\text{B}t / \hbar} \hat{F}_\text{AD} \hat{\rho}^\text{eq}_\text{B}] \notag\\
    &= e^{i (E_D - E_A) t / \hbar} \times C_{ff}(t).
\end{align}
\end{widetext}
Note that in the fourth line of the above Eq.~\ref{eq:CDA-Cff}, we performed terms substitution according to
\begin{subequations}
\begin{align}
    \hat{F}_\text{DA} &= e^{- \hat{S}_D} \hat{H}_{DA} e^{\hat{S}_A}, \label{eq:FDA-2} \\
    \hat{F}_\text{AD} &= e^{- \hat{S}_A} \hat{H}_{AD} e^{\hat{S}_D}, \label{eq:FAD-2} 
\end{align}
\end{subequations}
where $\hat{F}_\text{DA}$ and $\hat{F}_\text{AD}$ has been defined previously in Eq.~\ref{eq:FDA} and \ref{eq:FAD}, respectively. The proof is provided in Supplementary Material, Section VII. 

Eq.~\ref{eq:CDA-Cff} has shown that $C_{DA}(t) = e^{i (E_D - E_A) t / \hbar} \times C_{ff}(t)$, thus the FGR rate expression in  Eq.~\ref{eq:FGR-NCD} is exactly the same as the FGR in Eq.~\ref{eq:FGR-LRT}. 
Note that the two FGR formalisms are given under different representations of the Hamiltonian, {\it i.e.}, Eq.~\ref{eq:FGR-LRT} uses the polaron transformed Hamiltonian, while Eq.~\ref{eq:FGR-NCD} uses the LVC Hamiltonian and does not explicitly carry out PT. Nevertheless, PT is embedded in the correlation function of Eq.~\ref{eq:CDA}.

\section{Computational details} \label{apdx:comp}
\subsection{Phonon bath spectral density and discretization} \label{apdx:comp-1}
The phonon bath spectral density is taken as the Ohmic form with exponential cutoff function, 
\begin{align} \label{eq:Ohmic}
    J_\text{vib}(\tilde{\omega})  = \frac{\pi}{2} \alpha \tilde{\omega} e^{- \tilde{\omega} / \omega_\mathrm{c}},
\end{align}
where $\alpha = 2 E_R / (\hbar \omega_\text{c})$ is the Kondo parameter, and $\omega_\mathrm{c}$ is the bath characteristic frequency. Here we take $\hbar \omega_\mathrm{c} = 20$ cm$^{-1}$ in all calculations unless specified (see Appendix~\ref{apdx:wc-dep-EGL}), so that $k_\text{B} T \gg \hbar \omega_\mathrm{c}$ for $T = 300$ K. 
The continuous spectral density is efficiently discretized using the strategy as follows~\cite{Wang_Thoss_2007},
\begin{subequations}
\begin{align}
    &\nu_j = - \omega_\mathrm{c} \ln [1 - j / (1 + N_b)], \\
    &\lambda_j = \sqrt{\frac{\alpha \omega_\mathrm{c} \nu_j}{2 (N_b + 1)}},
\end{align}
\end{subequations}
where $j = 1, \cdots, N_b$, and $N_b$ is the number of bath oscillators. Here we take $N_b = 100$ in all calculations. 

\subsection{Effective spectral density and discretization} \label{apdx:comp-3}
With the presence of cavity loss, we use the correlation function in Eq.~\ref{eq:hgf_t_loss} to evaluate the FGR rate constants, where the $\{\omega_k, c_k\}$ parameters are sampled from the effective spectral density function in Eq.~\ref{eq:jeff} using the following ``equal frequency'' strategy~\cite{Burghardt_2009_1, Burghardt_2009_2, Hu_JPCL2023},
\begin{subequations} \label{eq:Jeff_sample_1}
\begin{align}
    \omega_k &= k\Delta \omega, \\
    c_k &= \sqrt{\frac{2}{\pi} J_\mathrm{eff}(\omega_k) \cdot \omega_k \cdot \Delta \omega},
\end{align}
\end{subequations}
where $k = 0, \cdots, N_c - 1$ and $N_c$ is the number of normal modes. Here, we use $N_c = 300$ in all calculations. Furthermore, $\Delta \omega = \omega_\text{max} / N_c$ is the normal mode frequency spacing, and $\omega_\text{max}$ is the cutoff frequency, which is properly chosen so that a satisfactory detailed sampling of the peak of $J_\mathrm{eff}(\tilde{\omega})$ is captured (and should recover the reorganization energy $\Lambda_\text{eff}$ according to Eq.~\ref{eq:reorg-eff}), and ensure the Poincar\'e recurrence time $2\pi / \Delta \omega$ is much longer than the characteristic time scale of the correlation functions. 

On the other hand, under the overdamped limit ($\mathcal{Q} \to 0$), the Brownian spectral density will reduce to the Drude-Lorentz one. By defining the characteristic frequency $\tilde{\omega}_\text{c} = \omega^2 / \Gamma$, Eq.~\ref{eq:jeff} reduces to
\begin{align} \label{eq:jeff-Drude}
    J_\text{eff}(\tilde{\omega}) = \frac{2\Lambda_\text{eff} \tilde{\omega}_\text{c} \tilde{\omega}}{\tilde{\omega}^2 + \tilde{\omega}^2_\text{c}}, 
\end{align}
where $\Lambda_\text{eff}$ is given in Eq.~\ref{eq:reorg-eff}. 
The ``equal frequency'' strategy in Eq.~\ref{eq:Jeff_sample_1} is no longer efficient to sample Eq.~\ref{eq:jeff-Drude} due to the long tailing of the Drude-Lorentz spectral density. For this case, we use the more efficient ``equal $\Lambda_\text{eff}$'' strategy instead, reading as~\cite{Wang_Thoss_2007}
\begin{subequations} \label{eq:Debye_discrete}
\begin{align}
    &\omega_k = \tilde{\omega}_\mathrm{c} \tan \left[ \frac{\pi}{2} \left(1 - \frac{k}{N_c + 1} \right) \right], \\
    &c_k = \sqrt{\frac{2 \Lambda_\text{eff}}{N_c + 1}} \omega_k,
\end{align}
\end{subequations}
where $k = 1, \cdots, N_c$ and $N_c = 300$ modes. 

In all realizations of sampling, convergence has been checked according to the reorganization energy, where
\begin{equation}
    \Big|1 - \hbar \omega \cdot \sum_k \frac{c^2_k}{\hbar \omega_k}\Big| < 1\%, 
\end{equation}
has been ensured. 

\subsection{The correlation function discretization and numerical fast Fourier transform (FFT)} \label{apdx:comp-4}
\begin{figure}[htbp]
    \centering
    \includegraphics[width=0.95\linewidth]{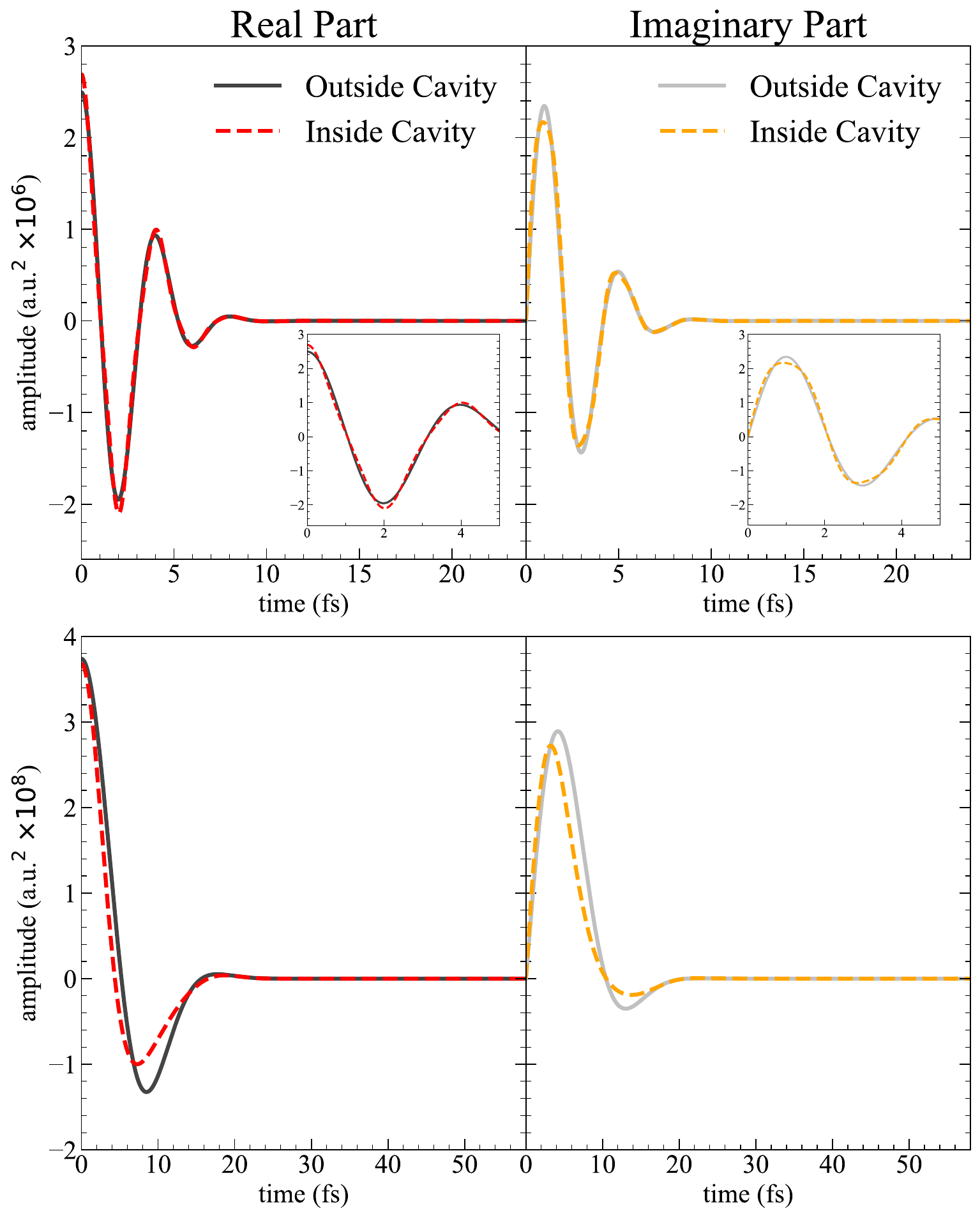}
    \caption{Visualization of typical correlation functions outside / inside the cavity. Upper panels: Model A (left for real part, right for imaginary part), where the short time behavior is also shown in the inner panels. Lower panels: Model B (left for real part, right for imaginary part). }
    \label{fig:5}
\end{figure}

To evaluate the FGR rate expression in Eq.~\ref{eq:FGR-LRT} under various donor-acceptor energy gaps ($- \Delta G_0$), we first discretize the correlation function $C_{ff}(t)$ (either for the lossless case with Eq.~\ref{eq:hgf_t} or the lossy case with Eq.~\ref{eq:hgf_t_loss}). The discretization strategy is as follows. For Model A, we choose a time step $dt = 2^{-19}$ fs, and the total time $t_\text{max} = 2^5$ fs in order to reach to convergence. For Model B, we choose a time step $dt = 2^{-16}$ fs, and the total time $t_\text{max} = 2^7$ fs. Visualization for typical correlation functions are shown in Fig.~\ref{fig:5}. The upper panels show the real part (left) and imaginary part (right) of the correlation functions outside (solid) and inside (dashed) the cavity using Model A parameters in Tabel~\ref{tab:par} and $T =$ 300 K. The inner panels zoom in the short time behaviors (from 0 to 5 fs), from which one sees that the dashed line fastly oscillates around the solid line. The lower panels are the same as the upper ones except for using Model B parameters in Tabel~\ref{tab:par} and $T =$ 300 K.

Then, we perform FFT to the discretized $C_{ff}(t)$ and take the real part to obtain the ET rate constants. FFT is performed using {\tt scipy.fftpack}. 

\subsection{Numerical evaluation of the GMJ expression in Eq.~\ref{eq:MJ_rate_loss_B}}
We aim to efficiently evaluate the $m$-fold discrete sum
\begin{align} \label{eq:Ialpha}
I_{m}(-\Delta G_0) &= \frac{1}{m!} \sum_{k_1,\dots,k_{m}} \Big(\prod_{\alpha=1}^{m} |g'_{k_\alpha}|^{\!2}\Big)\, \\
&\times \exp\!\left[- \beta\frac{\left(- \Delta G_0 - E_R - \sum_{\alpha=1}^{m} \hbar \omega_{k_\alpha}\right)^2}{4\,E_R}\right], \notag
\end{align}
on an equally spaced grid $\omega_k = k\,\Delta \omega$, where $k=0, \cdots, N_c-1$, and $N_c = 300$ (recall the sampling stragegy in Eq.~\ref{eq:Jeff_sample_1}). The kernel in Eq.~\ref{eq:Ialpha} depends only on the \emph{sum} $S = \sum_{\alpha=1}^{m} \hbar \omega_{k_\alpha}$, which allows us to reduce the $m$-fold sum to an $m$-fold linear self-convolution of a one-dimensional weight sequence, followed by a vectorized accumulation. Details are as follows.

Define $a_k \equiv |g'_k|^{2}$ and let
\begin{equation}
\label{eq:convalpha}
c^{(m)} \;\equiv\; \underbrace{a * a * \cdots * a}_{m\ \text{times}}
\end{equation}
be the $m$-fold \emph{linear} self-convolution of $a$. The resulting array has length
\begin{equation}
\label{eq:Lalpha}
    L_{m} \;=\; m\,N_c - (m - 1),
\end{equation}
and it naturally lives on the sum grid
\begin{equation}
\label{eq:Sgridalpha}
    S_s \;=\; s\,\Delta \omega,
\qquad s=0,1,\ldots,L_{m}-1.
\end{equation}
We compute $c^{(m)}$ efficiently in the frequency domain using FFT. Let $n_{\mathrm{fft}}\ge L_{m}$ (typically the next power of two), and for convenience we denote the real FFT and its inverse by $\mathrm{rFFT}$ and $\mathrm{iRFFT}$, respectively. 

Define $A \;=\; \mathrm{rFFT}\big(a,\, n_{\mathrm{fft}}\big)$, then 
\begin{equation} \label{eq:fftalpha}
    c^{(m)} \;=\; \mathrm{iRFFT}\!\big(A^{\,m},\, n_{\mathrm{fft}}\big)\big|_{0:L_{m}},
\end{equation}
which yields the \emph{linear} $m$-fold self-convolution with a computational complexity of $O(m\,N_c\log N_c)$, offering orders-of-magnitude speedups -- compared to the na\"{i}ve $O(m\, \exp{(m\, \log N_c}))$ if one directly performs the summation in Eq.~\ref{eq:Ialpha} with loops. 

Next, for a given query of $-\Delta G_0$ (vectorized, with dimension $\mathbb{R}^n$), define
\begin{equation}
\label{eq:udef}
u_{\ell} \;\equiv\; -\Delta G_{0,\ell} - E_R,
\end{equation}
where $-\Delta G_{0,\ell}$ denotes the $\ell$-th element of the $-\Delta G_0$ vector, with $\ell=1,\dots,n$. 
Then, assemble the Gaussian kernel on the outer-difference grid as follows,
\begin{equation}
\label{eq:kernelalpha}
    K_{\ell,s} \;=\; \exp\!\left[-\beta \frac{\big(u_{\ell} - S_s\big)^2}{4\,E_R}\right],
\end{equation}
with $s=0,\dots,L_{m}-1$. 
Combining Eqs.~\ref{eq:fftalpha} and \ref{eq:kernelalpha}, the summation in Eq.~\ref{eq:Ialpha} can be represented as a matrix–vector product for all entries of $-\Delta G_0$ at once:
\begin{equation}
\label{eq:finalalpha}
I_{m}(-\Delta G_{0,\ell}) \;=\; \frac{1}{m!} \sum_{s=0}^{L_{m}-1} c^{(m)}_{s}\,K_{\ell,s},
\end{equation}
or in vector form $I_{m}(-\Delta G_0) \;=\; (1 / m!)\, K\,c^{(m)}$, being highly numerical efficient. 

There are a few more details to mention. (i) Equally spaced grid ($\omega_k = k\,\Delta \omega$) is necessary to apply this method. (ii) All arrays are stored in \texttt{float64}.
(iii) To avoid division by zero in $g'_k$, we clip the grid as $\omega_k \leftarrow \max(\omega_k,\varepsilon)$ with $\varepsilon=10^{-15}$. 
(iv) The kernel assembly uses an outer difference, {\it e.g.}\ \verb|np.subtract.outer(u, S)|, to avoid broadcasting errors and Python-level loops. 
(v) When $m$ or $L_{m}$ is large, Eq.~\ref{eq:finalalpha} shall be evaluated in blocks along $-\Delta G_0$ to limit peak memory (although not necessary here for largest $m = 10$).

\section{Phonon bath characteristic frequency dependence of the energy gap law} \label{apdx:wc-dep-EGL}
According to the discussions in Section~\ref{sec:fast-cavity} and Section~\ref{sec:slow-cavity}, one expects that the EGL scaling relation ({\it i.e.}, the slope fitted from the $k_{\text{D}\to \text{A}}$ v.s. $-\Delta G_0$ diagram under a large donor-acceptor gap) will depend on the bath phonon characteristic frequency $\omega_\mathrm{c}$. For simplicity, we fix $T$ = 0. 

\begin{figure*}[htbp]
    \centering
    \includegraphics[width=0.8\linewidth]{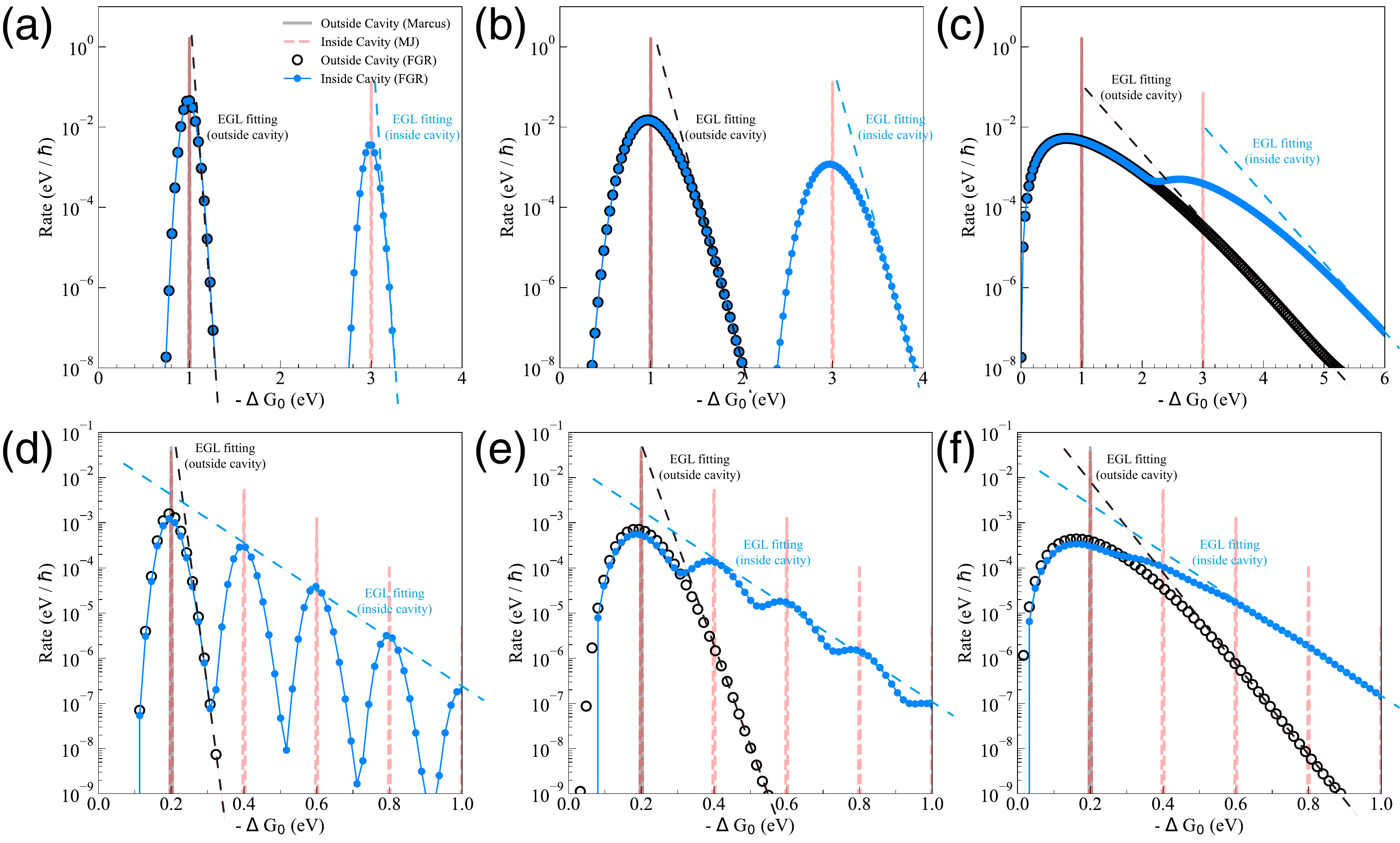}
    \caption{Effect of the bath phonon characteristic frequency to the EGL scaling relation. Panels (a)-(c) uses Model A parameters with $\hbar \omega_\mathrm{c}$ = 20 cm$^{-1}$, 200 cm$^{-1}$, and 2000 cm$^{-1}$, respectively. Results using Marcus (Eq.~\ref{eq:Marcus}) and MJ rate expression (Eq.~\ref{eq:MJ_rate_A}) are also presented (with $T$ = 0.01 K). 
    Panels (d)-(f) uses Model B parameters with $\hbar \omega_\mathrm{c}$ = 20 cm$^{-1}$, 100 cm$^{-1}$, and 300 cm$^{-1}$, respectively. Results using Marcus (Eq.~\ref{eq:Marcus}) and MJ rate expression (Eq.~\ref{eq:MJ_rate_B}) are also presented (with $T$ = 0.01 K).}
    \label{fig:6}
\end{figure*}

Fig.~\ref{fig:6}a-c presents the $k_{\text{D}\to \text{A}}$ v.s. $-\Delta G_0$ diagrams using the parameters of Model A with different bath phonon characteristic frequencies $\hbar \omega_\mathrm{c}$ = 20 cm$^{-1}$, 200 cm$^{-1}$, and 2000 cm$^{-1}$, respectively. 
One sees that the width of the peaks get broader as $\omega_\mathrm{c}$ increases. Furthermore, the slope of the fitted lines increase accordingly (to smaller negative values), while remain to be the same for outside the cavity (black dashed lines) and inside the cavity cases (blue dashed lines) -- as long as $\omega_\mathrm{c} < \omega$. This is in accordance with the theoretical prediction in Eqs.~\ref{eq:EGL_A_1}-\ref{eq:EGL_A_2}.  

Fig.~\ref{fig:6}d-f presents similar plots as Fig.~\ref{fig:6}a-c, but with Model B parameters and bath phonon characteristic frequencies $\hbar \omega_\mathrm{c}$ = 20 cm$^{-1}$, 100 cm$^{-1}$, and 300 cm$^{-1}$, respectively. One sees that as $\hbar \omega_\mathrm{c}$ increases, the outside cavity curve (black open circles) becomes broader and with an increasing slope; the inside cavity curves (blue dotted lines) also get broader but the slope remains unchanged (see blue dashed lines). This is because the cavity frequency $\omega$ remains unchanged and $\omega_\mathrm{c} < \omega$ always holds, being in accordance with the theoretical prediction in Eq.~\ref{eq:EGL_B_1}.

\section{Numerical performance of the generalized Marcus-Jortner rate expressions in Eqs.~\ref{eq:MJ_rate_loss_A} and \ref{eq:MJ_rate_loss_B}} \label{apdx:MJ_Rate_loss}

In this section, we further provide numerical demonstration for the GMJ expressions in Eqs.~\ref{eq:MJ_rate_loss_A} and \ref{eq:MJ_rate_loss_B} and compare them with the FGR results using Eq.~\ref{eq:FFCF-3}. We fix $T =$ 300 K so that the high-temperature limit for the phonon bath holds. 

\begin{figure*}
    \centering
    \includegraphics[width=0.8\linewidth]{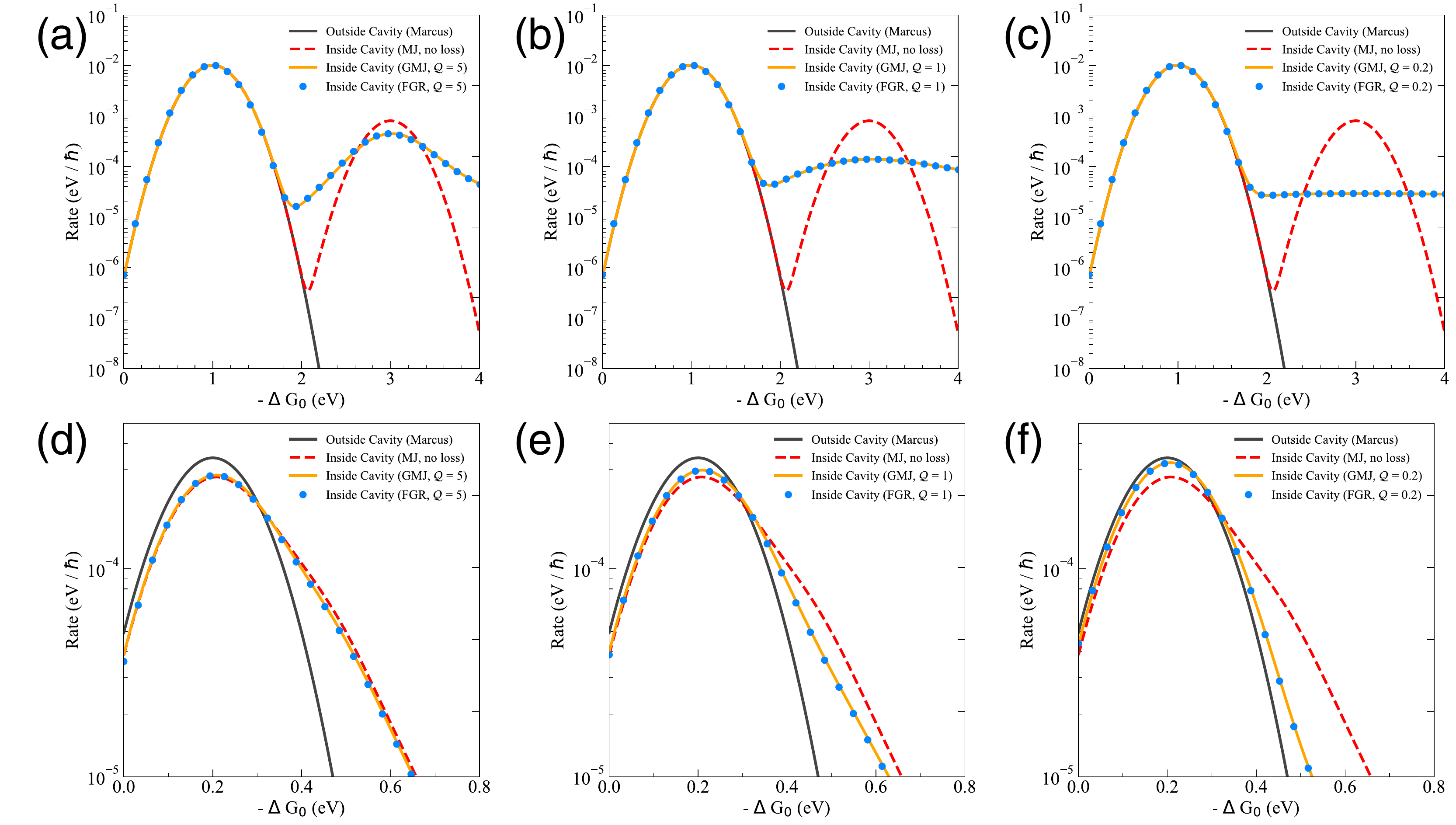}
    \caption{Comparison between the GMJ and FGR (with Eq.~\ref{eq:FFCF-3}) results for ET rates inside lossy cavities. The outside cavity Marcus rates (Eq.~\ref{eq:Marcus}, black solid lines) and inside a lossless cavity MJ rates (Eqs.~\ref{eq:MJ_rate_A} or \ref{eq:MJ_rate_B}, red dashed lines) are also presented for reference. Panels (a)-(c) uses Model A parameters with cavity $\mathcal{Q}$-factors 5, 1, and 0.2, respectively. The GMJ results are obtained using Eq.~\ref{eq:MJ_rate_loss_A}. 
    Panels (d)-(f) uses Model B parameters with cavity $\mathcal{Q}$-factors 5, 1, and 0.2, respectively. The GMJ results are obtained using  Eq.~\ref{eq:MJ_rate_loss_B}.  }
    \label{fig:7}
\end{figure*}

Fig.~\ref{fig:7}a-c shows the $k_{\text{D}\to \text{A}}$ v.s. $-\Delta G_0$ plots using Model A parameters, where $\mathcal{Q} =$ 5, 1, 0.2, respectively. For comparison, the outside cavity Marcus rate (Eq.~\ref{eq:Marcus}, black solid line) and inside the cavity without loss (Eq.~\ref{eq:MJ_rate_A}, red dashed line) results are also presented.
One sees that as $\mathcal{Q}$ decreases, the peak centered at $- \Delta G_0 =$ 3 eV gradually disappear. And the GMJ results (Eq.~\ref{eq:MJ_rate_loss_A}, orange solid line) precisely agree with the FGR results (Eq.~\ref{eq:FFCF-3}, blue dots) across all the parameter regime explored. 

Fig.~\ref{fig:7}d-f shows similar plots as Fig.~\ref{fig:7}a-c but with Model B parameters. One sees that as $\mathcal{Q}$ decreases, the orange curve asymptotically reduce to the black solid curve (outside cavity Marcus rate). And still, the GMJ theory (Eq.~\ref{eq:MJ_rate_loss_B}, orange solid line) agrees well with the FGR rates (Eq.~\ref{eq:FFCF-3}, blue dots) across all the parameter regimes explored. 

\section*{References}
\bibliography{ref}

\end{document}